\documentclass{iau-JDSS}
\usepackage{graphicx}

\def\ltsima{$\; \buildrel < \over \sim \;$}
\def\simlt{\lower.5ex\hbox{\ltsima}}
\def\gtsima{$\; \buildrel > \over \sim \;$}
\def\simgt{\lower.5ex\hbox{\gtsima}}

\def\arcsec{$^{\prime\prime}$}

\title[$Herschel$ View of Star Formation] 
{The $Herschel$ View of Star Formation}

\author[Ph. Andr\'e]   
{Philippe Andr\'e}

\affiliation{$^1$Laboratoire d'Astrophysique (AIM) Paris-Saclay, 
CEA Saclay, 91191 Gif-sur-Yvette, France \break email: pandre@cea.fr}

\pubyear{2013}
\volume{Volume 16}  
\pagerange{}
\date{?? and in revised form ??}
\jname{Highlights of Astronomy, Volume 16}
\editors{Thierry Montmerle, ed.}
\begin{document}

\maketitle

\begin{abstract}
Recent studies of the nearest star-forming clouds of the Galaxy at submillimeter wavelengths with the $Herschel$ Space Observatory have provided us with unprecedented images of the initial conditions and early phases of the star formation process. 
The $Herschel$ images reveal an intricate network of filamentary structure in every interstellar cloud. These filaments all exhibit remarkably similar widths - about a tenth of a parsec - but only the densest ones contain prestellar cores, the seeds of future stars. The $Herschel$ results favor a scenario in which interstellar filaments and prestellar cores represent two key steps in the star formation process: first turbulence stirs up the gas, giving rise to a universal web-like structure in the interstellar medium, then gravity takes over and controls the further fragmentation of filaments into prestellar cores and ultimately protostars. This scenario provides new insight into the inefficiency of star formation, the origin of stellar masses, and the global rate of star formation in galaxies. Despite an apparent complexity, global star formation may be governed by relatively simple universal laws from filament to galactic scales.
\keywords{stars: formation -- ISM: clouds -- ISM: Filaments --  ISM: structure -- submillimeter}
\end{abstract}

\firstsection 
\section{Introduction \label{intro}}

Star formation is one of the most complex processes in astrophysics, involving a subtle interplay between gravity, turbulence, 
magnetic fields, feedback mechanisms, heating and cooling effects etc...
Yet, despite this apparent complexity, the net products of the star formation process on global scales are relatively simple and robust.
In particular, the distribution of stellar masses at birth or stellar initial mass function (IMF) is known to be quasi-universal 
(e.g. Kroupa 2001, Chabrier 2005, Bastian et al. 2010).
Likewise, the star formation rate 
on both GMC and galaxy-wide scales is related to the mass of (dense molecular) gas available 
by rather well defined ``star formation laws''  (e.g. Kennicutt 1998, Gao \& Solomon 2004, Lada et al. 2010). 
On the basis of recent submillimeter imaging observations 
obtained with the $Herschel$ Space Observatory (Pilbratt et al. 2010) on Galactic interstellar clouds
as part of the Gould Belt (Andr\'e et al. 2010), HOBYS (Motte et al. 2010), and Hi-GAL (Molinari et al. 2010) surveys, 
the thesis advocated in this paper 
is that it may be possible to explain, at least partly, the IMF and the global rate of star formation 
in terms of the quasi-universal filamentary structure of the cold interstellar medium (ISM) out of which stars form.

In particular, the bulk of nearby ($d \simlt 500$~pc) molecular clouds, 
mostly located in Gould's Belt (e.g. Guillout 2001, Perrot \& Grenier 2003), have been imaged 
at 6 wavelengths between 70~$\mu$m and 500~$\mu$m as part of  
the $Herschel$ Gould Belt survey (HGBS -- Andr\'e et al. 2010).  
Observationally, the molecular clouds of the Gould Belt are the best laboratories at our disposal to investigate the star 
formation process in detail, at least as far as solar-type stars are concerned. 
The $\sim 15$  nearby clouds covered by the HGBS  span a wide range of 
environmental conditions, from 
active, cluster-forming complexes 
such as the Orion A \& B GMCs or the Aquila Rift cloud complex (e.g. Gutermuth et al. 2008) to quiescent regions 
with no star formation activity whatsoever such as the Polaris flare {\it translucent} cloud (e.g. Heithausen et al. 2002). 
The main scientific goals of the HGBS are to clarify the nature of the relationship between 
the prestellar core mass function (CMF) and the stellar IMF (cf. \S ~3 below) and to elucidate the physical mechanisms responsible 
for the formation of prestellar cores out of the diffuse ISM (cf. \S ~5 and \S ~7). 

This paper presents an overview of the first results obtained with $Herschel$ on nearby star-forming clouds. 
Section~2 emphasizes the universality of the filamentary structure revealed by $Herschel$ in the cold ISM.
Section~3 presents preliminary results obtained on the global properties of prestellar dense cores. 
Section~4 summarizes a few theoretical considerations on the gravitational instability of filamentary clouds. 
Section~5 presents the observational evidence of a column density threshold for the formation of prestellar cores 
and shows how this can be interpreted in terms of the gravitational instability threshold of interstellar filaments. 
Section~6 discusses implications of the $Herschel$ results on filaments and cores for our understanding 
of the origin of the IMF and the global rate of star formation in galaxies. 
Finally, Sect.~7 concludes by summarizing the scenario of star formation emerging from the $Herschel$ results.

\begin{figure}[!h]
\center
\hspace{-3mm}
  \resizebox{6.75cm}{!}{     
\includegraphics[angle=0,scale=0.47]{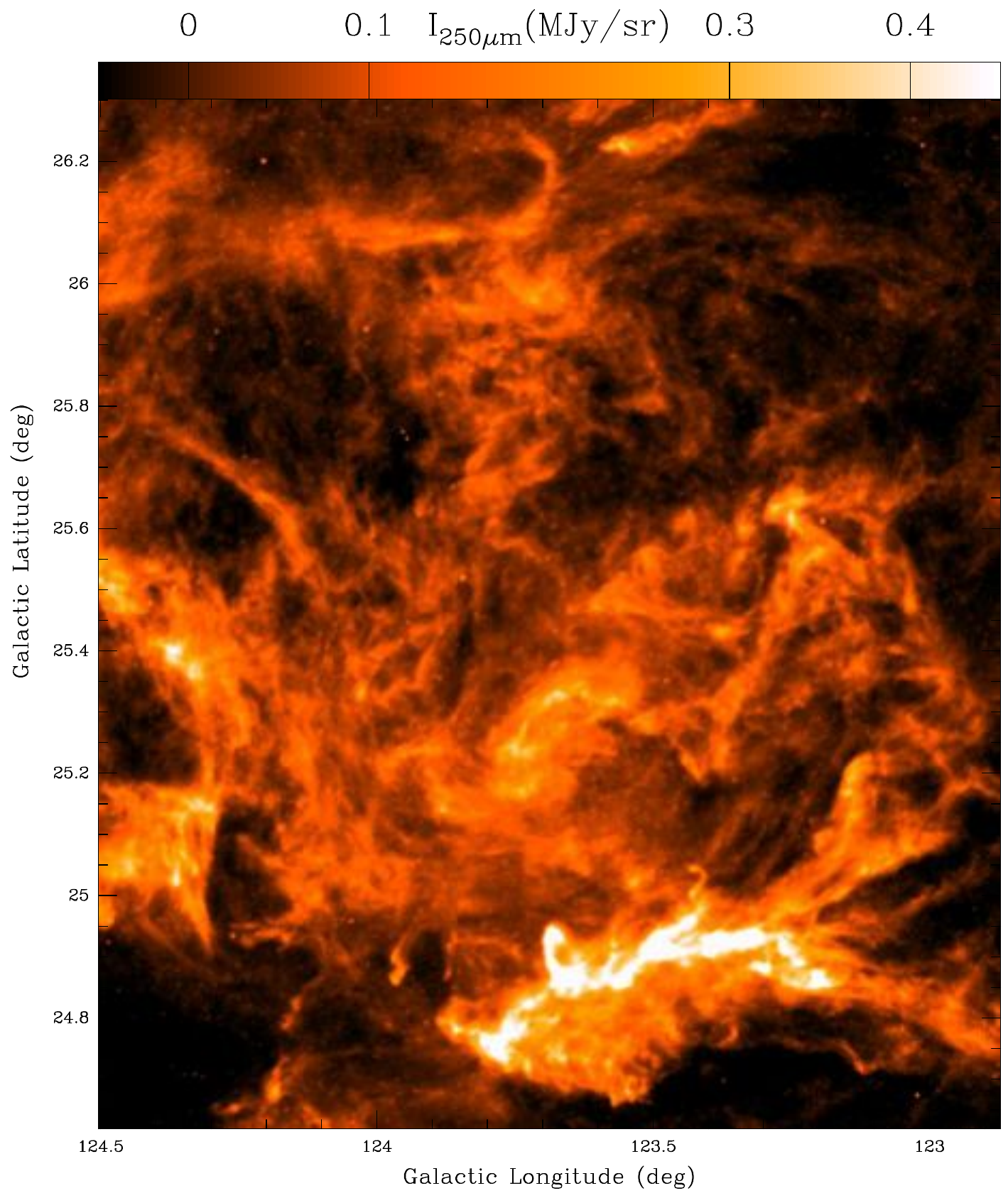}}      
 \hspace{-4mm}
  \resizebox{7.15cm}{!}{
 \includegraphics[angle=0,scale=0.5]{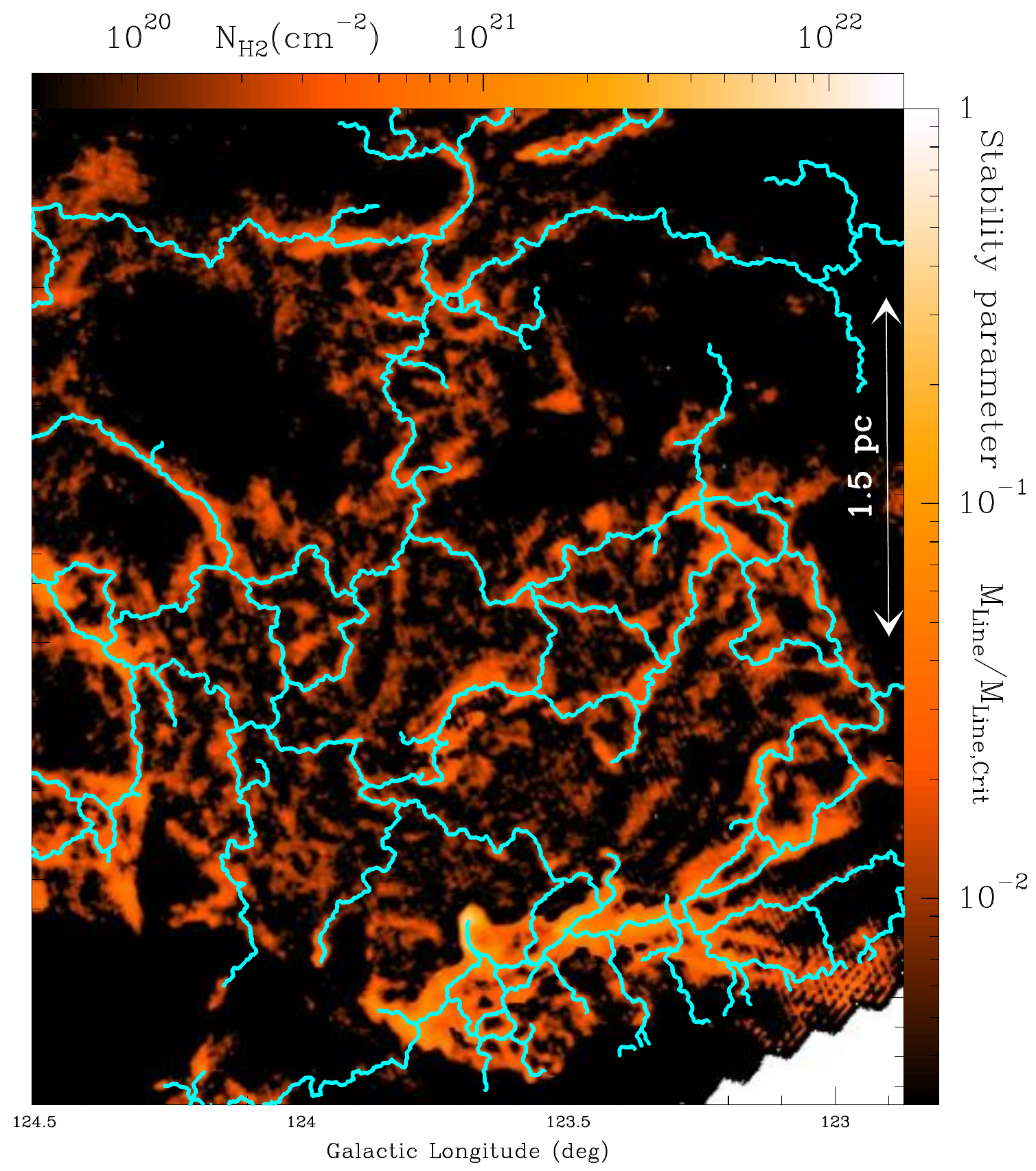}}
\caption{
{\bf Left:} $Herschel$/SPIRE 250~$\mu$m dust continuum map of a portion of the Polaris flare translucent cloud 
($d \sim 150$~pc) taken as part of the HGBS survey (e.g. Miville-Desch\^enes et al. 2010, Ward-Thompson et al. 2010).
{\bf Right:} Corresponding column density map derived from  $Herschel$ data (Andr\'e et al. 2010). 
The contrast of the filaments has been enhanced using a curvelet transform (cf. Starck et al. 2003). 
The skeleton of the filament network identified 
with the DisPerSE algorithm (Sousbie 2011) is shown in light blue. 
Given the typical filament width $\sim $~0.1~pc  (Arzoumanian et al. 2011 -- see Fig.~4 below), this column density map 
is equivalent to a {\it map of the mass per unit length along the filaments} (see color scale on the right). 
\label{aquila_polaris_filaments}
}
\end{figure}

\section{Universality of the filamentary structure in the cold ISM}

The high quality and dynamic range of the $Herschel$ images are such that they provide key information on 
the structure of molecular clouds on a wide range of spatial scales from the size of entire cloud complexes  ($\simgt 10$~pc)
down to the scale of individual dense cores ($< 0.1$~pc). 
In particular, one of the most spectacular early findings made with $Herschel$
is the omnipresence of long ($> \rm{pc} $ scale) filamentary structures in 
the cold ISM and the apparently tight connection between the filaments and the 
formation process of dense cloud cores (e.g. Andr\'e et al. 2010; Men'shchikov et al. 2010; Molinari et al. 2010). 
While interstellar clouds were already known to exhibit large-scale filamentary 
structures long before $Herschel$ (e.g. Schneider \& Elmegreen 1979; 
Abergel et al. 1994; 
Hartmann 2002; Hatchell et al. 2005; Myers 2009), $Herschel$ now demonstrates 
that these filaments are truly ubiquitous in the giant molecular clouds (GMCs) of our Galaxy
(Molinari et al. 2010) and provides 
an unprecedented large-scale view of the role of filaments in the formation of prestellar cores (see \S ~5 below). 
Filaments are omnipresent even in diffuse, non-star-forming complexes such as the Polaris translucent cloud 
(cf. Fig.~1 -- Miville-Desch\^enes et al. 2010; Ward-Thompson et al. 2010), suggesting that 
the formation of filaments {\it precedes} star formation in the cold ISM. 
Importantly, the few high-resolution spectral line observations 
available to date suggest that the filaments seen in the $Herschel$ dust continuum images are velocity-coherent structures 
(e.g. Hacar \& Tafalla 2011; Li \& Goldsmith 2012; Arzoumanian et al. 2013).

A very common filamentary pattern observed with $Herschel$ is that of a main filament surrounded by a population of 
fainter ``sub-filaments'' or striations approaching the main filament from the side and apparently connected to it 
(see Fig.~2 and Palmeirim et al. 2013, Peretto et al. 2012, Hennemann et al. 2012, Cox et al. 2013). 
The morphology of these ``sub-filaments'' and striations is suggestive of accretion flows feeding the main filaments 
with surrounding cloud material.

  \begin{figure*}[!h]
   \vspace{-5pt}
   \centering
     \resizebox{\hsize}{!}{     
    \includegraphics[angle=0]{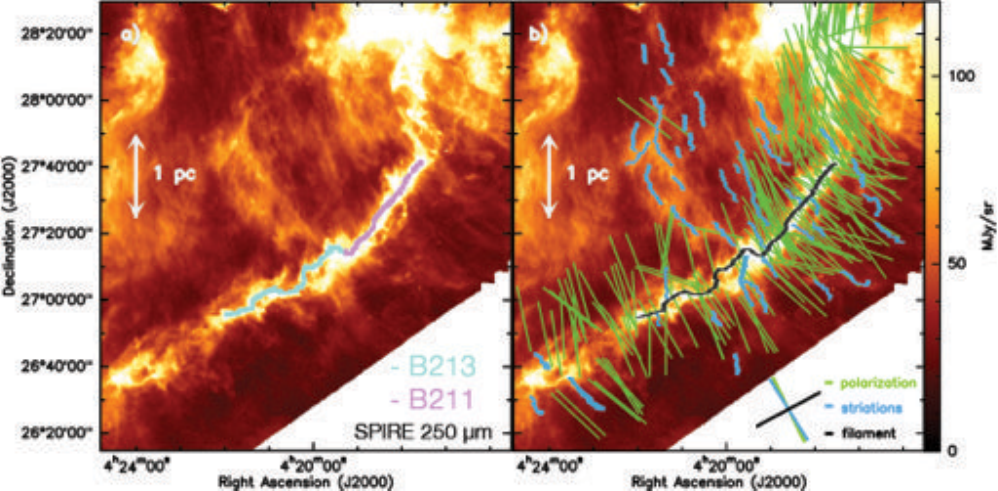}}  
   \caption{ {\bf(a)} $Herschel$/SPIRE 250 $\mu$m dust continuum image of the B211/B213/L1495 region in Taurus  ($d \sim 140$~pc). 
   The light blue and purple curves show the crest of the B213/B211 filament.
    {\bf(b)} Display of optical and infrared polarization vectors from (e.g. Heyer et al. 2008; Chapman et al. 2011)
    tracing the magnetic field orientation in the same region, overlaid on the $Herschel$/SPIRE 250 $\mu$m image. 
   The plane-of-the-sky projection of the magnetic field appears to be oriented perpendicular to the B211/B213 filament and roughly aligned with the 
   general direction of the striations overlaid in blue.  (From Palmeirim et al. 2013.)
  	}
              \label{pol}
   \end{figure*}

More generally, in any given cloud complex, $Herschel$ imaging reveals a whole network of filaments (see Fig.~1), making 
it possible to characterize their properties in a statistical manner. Furthermore, the $Herschel$ maps have resolved the structure 
of nearby filaments with unprecedented detail. Detailed analysis of the radial column density profiles 
derived from $Herschel$ data suggests that the shape of the filament radial  profiles is quasi universal and 
well described by a Plummer-like function of the form (cf. Palmeirim et al. 2013 and Fig.~3):
$$ \rho_{p}(r) = \frac{\rho_{c}}{\left[1+\left({r/R_{\rm \rm flat}}\right)^{2}\right]^{p/2}}\  \longrightarrow 
 \Sigma_{p}(r) = A_{p}\,  \frac{\rho_{\rm c}R_{\rm \rm flat}}{\left[1+\left({r/R_{\rm \rm flat}}\right)^{2}\right]^{\frac{p-1}{2}}}, \ \ (1)$$

\noindent 
where $\rho_{c}$ is the central density of the filament, $R_{\rm \rm flat}$ is the radius of the flat inner region, $p \approx 2$ 
is the power-law exponent at large radii ($r$$>>$$R_{\rm flat}$), $A_p $
is a finite constant factor 
which includes the effect of the filament's inclination angle to the plane of the sky.
Note that the density structure of an isothermal gas cylinder in hydrostatic equilibrium follows Eq.~(1) with $p = 4$ (Ostriker 1964), 
instead of the observed $p \approx 2$ value.

 \begin{figure*}[!h]
      \hspace{-6mm}
  \resizebox{7.5cm}{!}{        
\includegraphics[angle=0]{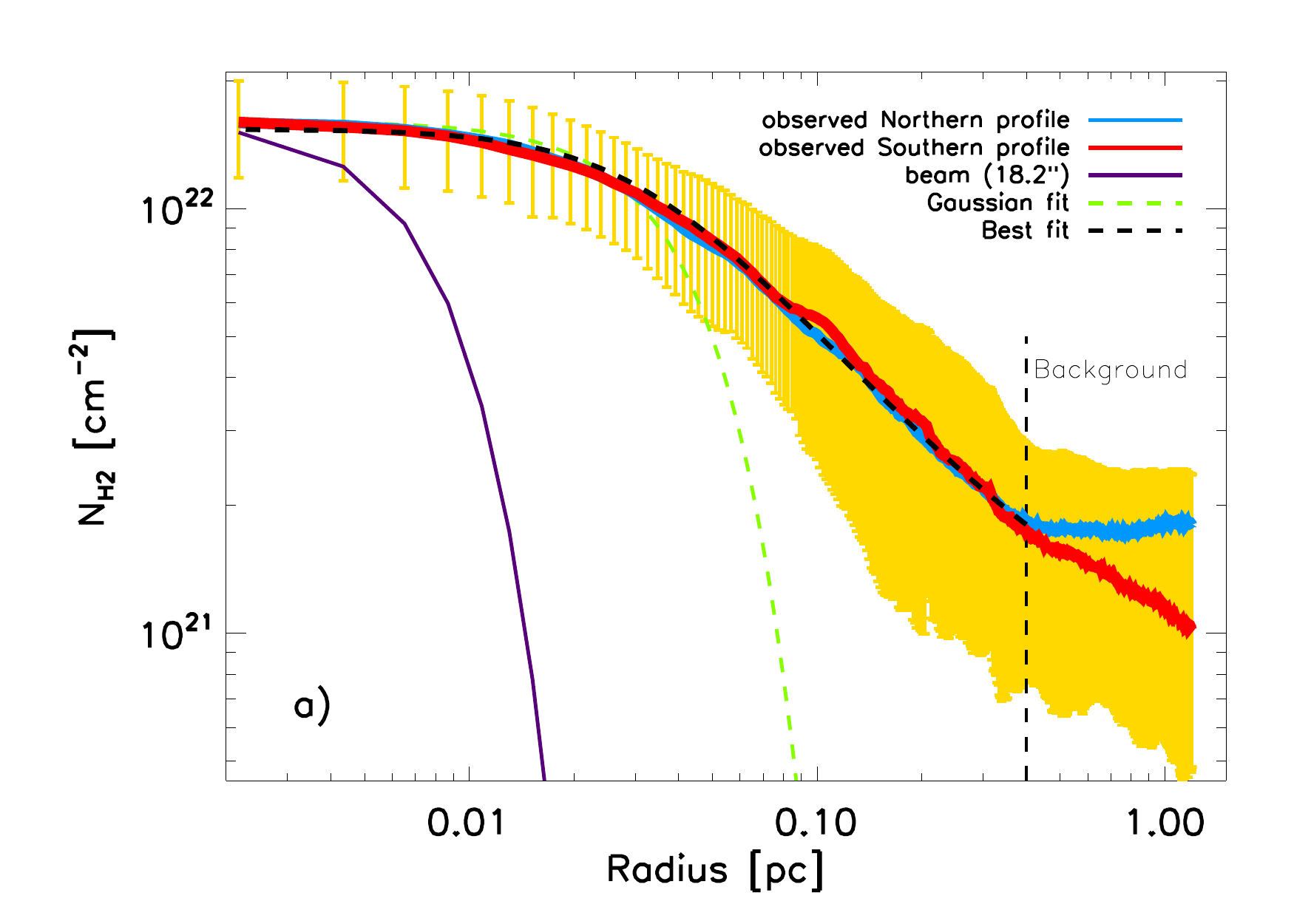}}  
      \hspace{-8mm}
  \resizebox{7.5cm}{!}{
 \includegraphics[angle=0]{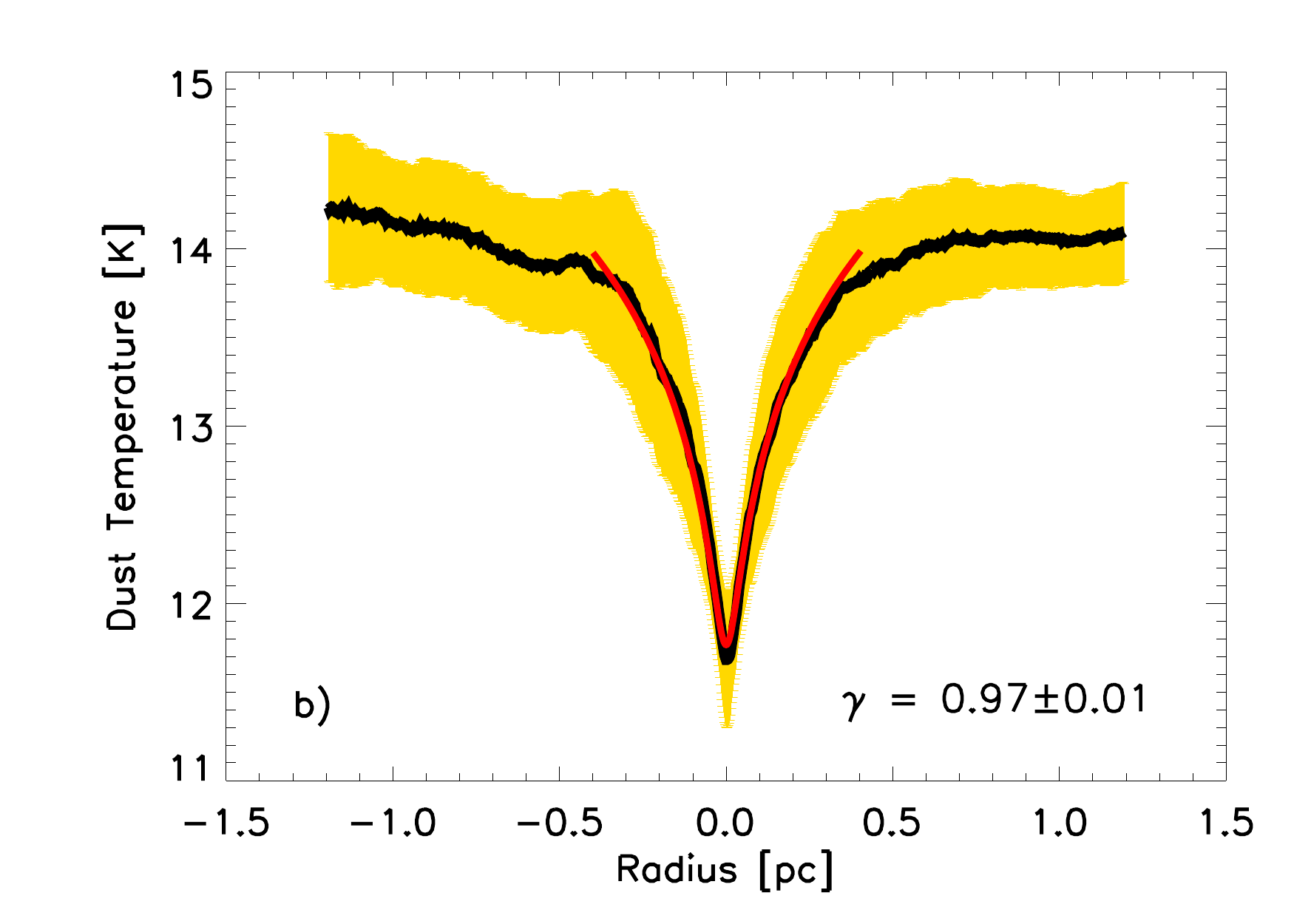}} 
   \caption{{\bf(a)} Mean radial column density profile observed with $Herschel$ 
perpendicular to the B213/B211 filament in Taurus (cf. Fig.~2), 
for both the Northern (blue curve) and the Southern part (red curve) of the filament.
The yellow area shows the ($\pm 1\sigma$) dispersion of the distribution of radial profiles along the filament.
The inner solid purple curve shows the effective 18\arcsec ~HPBW resolution
(0.012~pc at 140~pc) of  the column density map used to construct the profile. 
The dashed black curve shows the best-fit Plummer model (convolved with the 18\arcsec ~beam) described by Eq.~(1) 
with $p$=2.0$\pm$0.4 and a diameter $2\times R_{\rm \rm flat} = 0.07 \pm 0.01$~pc, which matches the data very well for $r$$\leq$0.4\,pc, 
{\bf(b)} Mean dust temperature profile measured perpendicular to the B213/B211 filament.
The solid red curve shows the best model temperature profile obtained by assuming that the filament has a density profile given 
by the Plummer model shown in {\bf(a)} (with $p = 2$) and obeys a polytropic equation of state, $P \propto \rho^{\gamma}$, and 
thus $T(r) \propto \rho(r)^{(\gamma-1)}$. This best fit corresponds to a polytropic index  $\gamma$=0.97$\pm$0.01. 
(From Palmeirim et al. 2013.)
 }
              \label{fil_prof}
    \end{figure*}

Remarkably,  the diameter $2 \times R_{\rm flat}$ of the flat inner plateau in the filament radial profiles appears to be roughly constant 
$\sim 0.1$~pc for all filaments,  at least in the nearby clouds of Gould's Belt  (cf. Arzoumanian et al. 2011). 
This is illustrated in Fig.~4 which shows that nearby interstellar filaments are characterized by a very narrow distribution of inner 
FWHM widths centered at about 0.1~pc.
 
The origin of this quasi-universal inner width of interstellar filaments is not yet well understood.
A possible interpretation  
is that it corresponds to the sonic scale below which interstellar turbulence becomes subsonic in diffuse, non-star-forming molecular gas 
(cf. Padoan et al. 2001 and \S ~7 below). In this view, the observed filaments would 
correspond to dense, post-shock stagnation gas associated with shocked-compressed regions resulting from converging flows
in supersonic interstellar turbulence. 
Interestingly, the filament width $\sim 0.1$~pc is also comparable to the cutoff wavelength 
$\lambda_A \sim 0.1\, \rm{pc} \times (\frac{B}{10\mu G}) \times (\frac{n_{H_2}}{10^3 \rm{cm}^{-3}})^{-1}$ 
for MHD waves in (low-density, primarily neutral) molecular clouds (cf. Mouschovias 1991), 
if the typical magnetic field strength is $B \sim 10\mu$G (e.g. Crutcher 2012). 
Alternatively, the characteristic width may also be understood if interstellar filaments are formed as quasi-equilibrium structures 
in pressure balance with a typical ambient ISM pressure $P_{\rm ext} {\sim} 2$$-$5$\times$$10^4 \, \rm{K\, cm}^{-3} $ 
(Fischera \& Martin 2012; Inutsuka et al., in prep.).
     
\begin{figure}
\center
\includegraphics[scale=1.0]{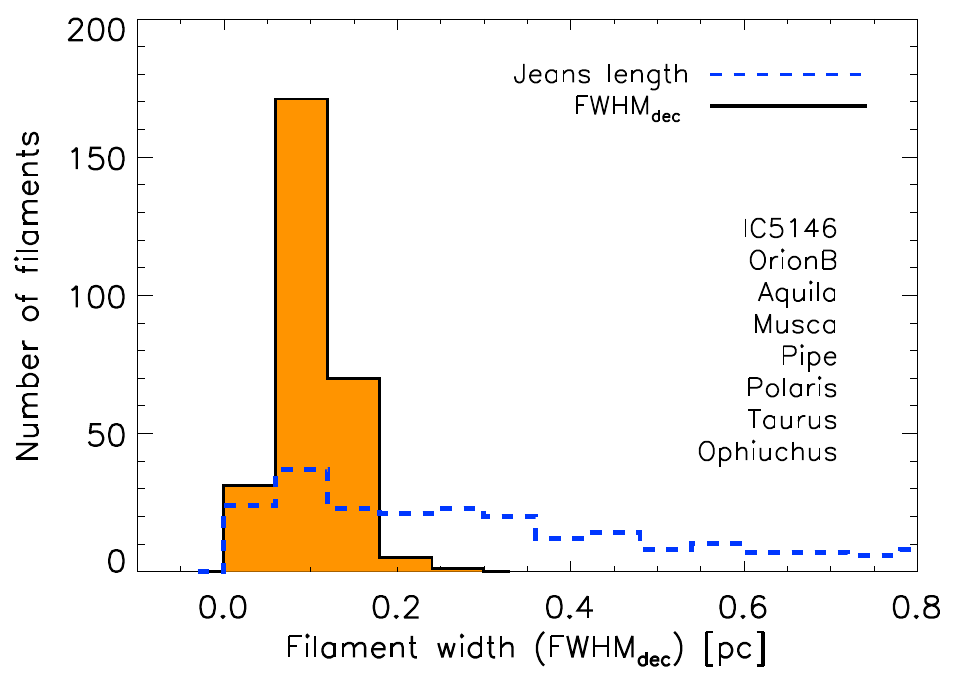} 
\caption{Histogram of deconvolved FWHM widths for a sample of 278 filaments in 8 nearby regions of the Gould Belt, 
all observed with $Herschel$ (at effective spatial resolutions ranging from $\sim 0.01$~pc to $\sim 0.04$~pc) 
and analyzed in the same way. The distribution of filament widths is narrow 
with a median value of 0.09~pc and a standard deviation of 0.04~pc. In contrast, the distribution of Jeans lengths corresponding 
to the central column densities of the filaments (blue dashed histogram) is much broader.  (Adapted from Arzoumanian et al. 2011.)
\label{background_ext}
}
\end{figure}

\begin{figure}
\center
\includegraphics[scale=0.55]{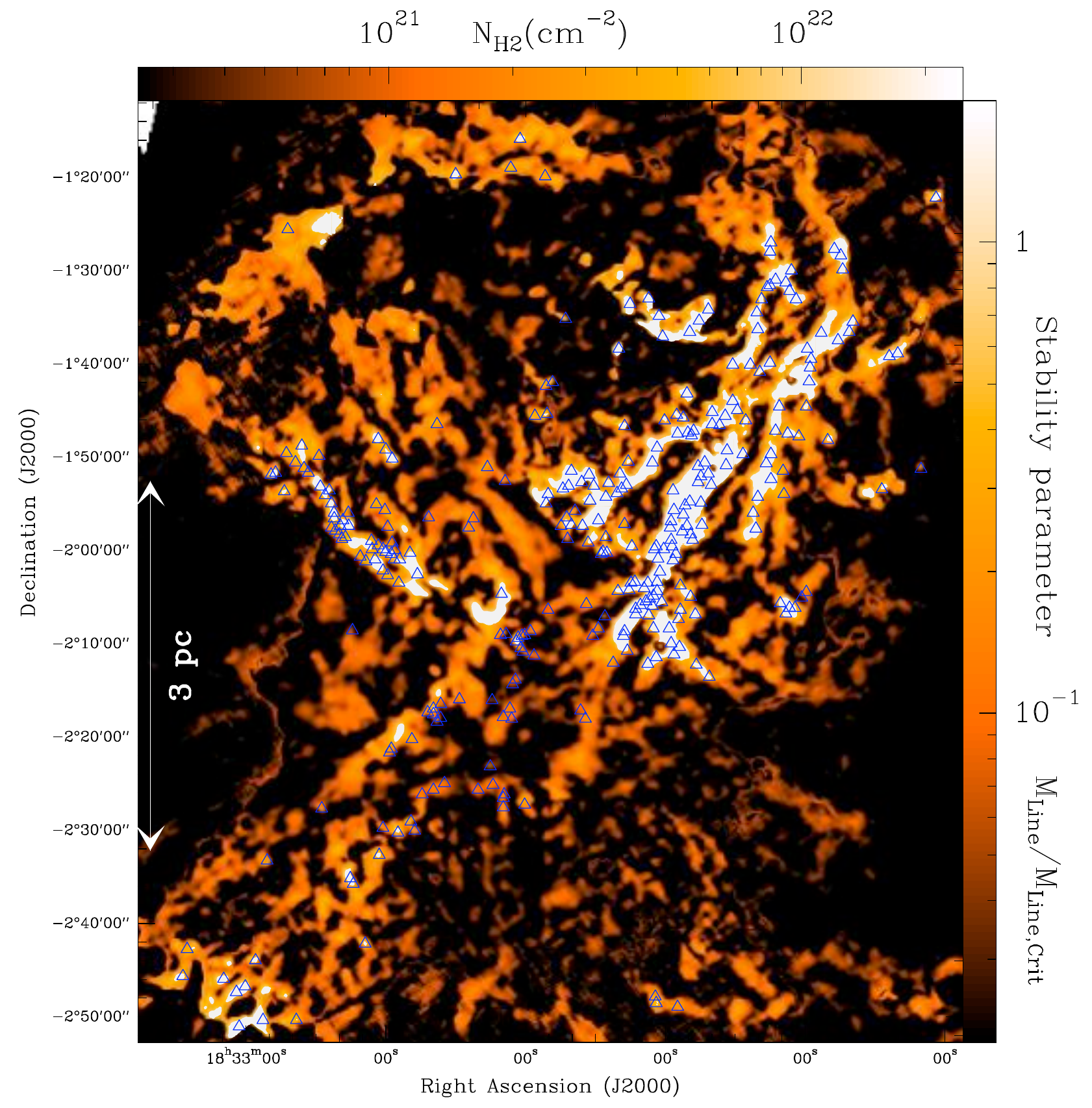}
\caption{
Column density map of a subfield of the Aquila star-forming region ($d \sim 260$~pc) 
derived from $Herschel$ data (Andr\'e et al. 2010). 
The contrast of the filaments has been enhanced using a curvelet transform (cf. Starck et al. 2003). 
Given the typical width $\sim $~0.1~pc of the filaments  (Arzoumanian et al. 2011 -- see Fig.~4), this  
map is  equivalent to a {\it map of the mass per unit length along the filaments}. 
The areas where the filaments have a mass per unit length larger than half the critical value $2\, c_s^2/G$ (cf. Inutsuka \& Miyama 1997 and \S ~4) 
and are thus likely gravitationally unstable have been highlighted in white. 
The bound prestellar cores identified by K\"onyves et al. (2010) in Aquila are shown as small blue triangles.
\label{aquila_polaris_filaments}
}
\end{figure}

\section{Dense cores and their ensemble properties as derived from  \emph{Herschel} observations}

As prestellar cores and deeply embedded protostars emit the bulk of their luminosity at far-infrared and submillimeter wavelengths, 
$Herschel$ observations are also ideally suited for taking a sensitive census of such cold objects in 
nearby molecular cloud complexes. 
This is indeed one of the main observational objectives of the $Herschel$ Gould Belt survey (HGBS).

Conceptually, a dense core is an individual fragment or local overdensity  
which corresponds to a local minimum in the gravitational potential of a molecular cloud. 
A starless core is a dense core with no central protostellar object. 
A prestellar core may be defined as a dense core which is both starless and self-gravitating.
In other words, a prestellar core is a 
self-gravitating condensation of gas and dust within a molecular cloud 
which may potentially form an individual star (or system)  by gravitational collapse 
(e.g. Motte et al. 1998; Andr\'e et al. 2000; Di Francesco et al. 2007; Ward-Thompson et al. 2007). 
Known prestellar cores are observed at the bottom of the hierarchy of interstellar cloud structures and depart from Larson (1981)'s self-similar 
scaling relations. They are the smallest units of star formation (e.g. Bergin \& Tafalla 2007) 
and correspond to ``coherent'' regions of nearly constant and thermal velocity dispersion which do not 
obey Larson (1981)'s power-law linewidth vs. size relation (Myers 1983; Goodman et al. 1998; Andr\'e et al. 2007).
To first order, known prestellar cores have simple, convex (and not very elongated) shapes, and their density structure approaches that 
of Bonnor-Ebert isothermal spheroids bounded by the external pressure exerted by the background parent cloud (e.g. Johnstone et al. 2000, 
Alves et al. 2001). 
Apart from an unprecedented mapping speed, two key advantages of $Herschel$ broad-band imaging for prestellar core surveys  
are 1) that dust continuum emission is 
largely optically thin at far-infrared/submillimeter wavelengths and thus directly tracing column density, and 2) that the $\sim 18$\arcsec$\,$HPBW 
angular resolution of $Herschel$ at $\lambda = 250\, \mu $m, corresponding to $\sim 0.03$~pc at a distance $d = 350$ pc, 
is sufficient to resolve the typical Jeans length in nearby clouds, which is also the characteristic diameter expected for Bonnor-Ebert-like cores.

While in general the gravitational potential cannot be inferred 
from observations, it turns out to be directly related to the observable column density distribution for the post-shock, filamentary cloud layers 
produced by supersonic turbulence in numerical simulations of cloud evolution (Gong \& Ostriker 2011). 
In practical terms, this means that one can define a dense core 
as the immediate vicinity of a local maximum in observed column density maps such as the maps derived from $Herschel$ imaging 
(see Fig.~5 and Fig.~6a for examples). 
In more mathematical terms, the projection of a dense core onto the plane of the sky corresponds 
to the ``descending 2-manifold'' (cf. Sousbie 2011) associated to a local maximum in column density, i.e., the set of points connected to the 
maximum by integral lines 
following the gradient 
of the column density distribution. 

In practice, systematic source/core extraction in wide-field dust continuum images of highly structured molecular clouds is a complex problem 
which can be conveniently decomposed into two sub-tasks: 1) source/core detection, and 2) source/core measurement. 
In the presence of noise and background cloud fluctuations, the sub-task of detecting source/cores reduces to identifying statistically significant 
intensity/column density peaks based on the information provided by the finite-resolution far-infrared/submillimeter continuum image(s) 
observed with, e.g., $Herschel$. The main problem to be solved in the other sub-task of measuring detected sources/cores is to find the 
spatial extent or ``footprint'' of each source/core, corresponding to the ``descending 2-manifold'' 
of the ``mathematical'' definition given above.  The {\it getsources} method devised by Men'shchikov et al. (2010, 2012) and used to identify 
cores in the HGBS data is a new approach to these two sub-tasks which makes full use of the multi-scale, multi-wavelength nature of the 
source extraction problem in the case of $Herschel$ data. 
Once cores have been extracted from the maps, the $Herschel$ observations provide a very sensitive way of distinguishing between 
protostellar and starless cores based on the presence or absence of 70~$\mu$m emission. 
The 70~$\mu$m flux is indeed known to be a very good tracer of the internal luminosity of a protostar (e.g. Dunham et al. 2008), 
and $Herschel$ observations of nearby clouds have the sensitivity to detect ``first protostellar cores'' (cf. Pezzuto et al. 2012), 
the very first and lowest-luminosity ($\sim 0.01$--0.1$\, L_\odot $) stage of protostars (e.g. Larson 1969; Saigo \& Tomisaka 2011).

\begin{figure}
\center
\includegraphics[scale=0.575]{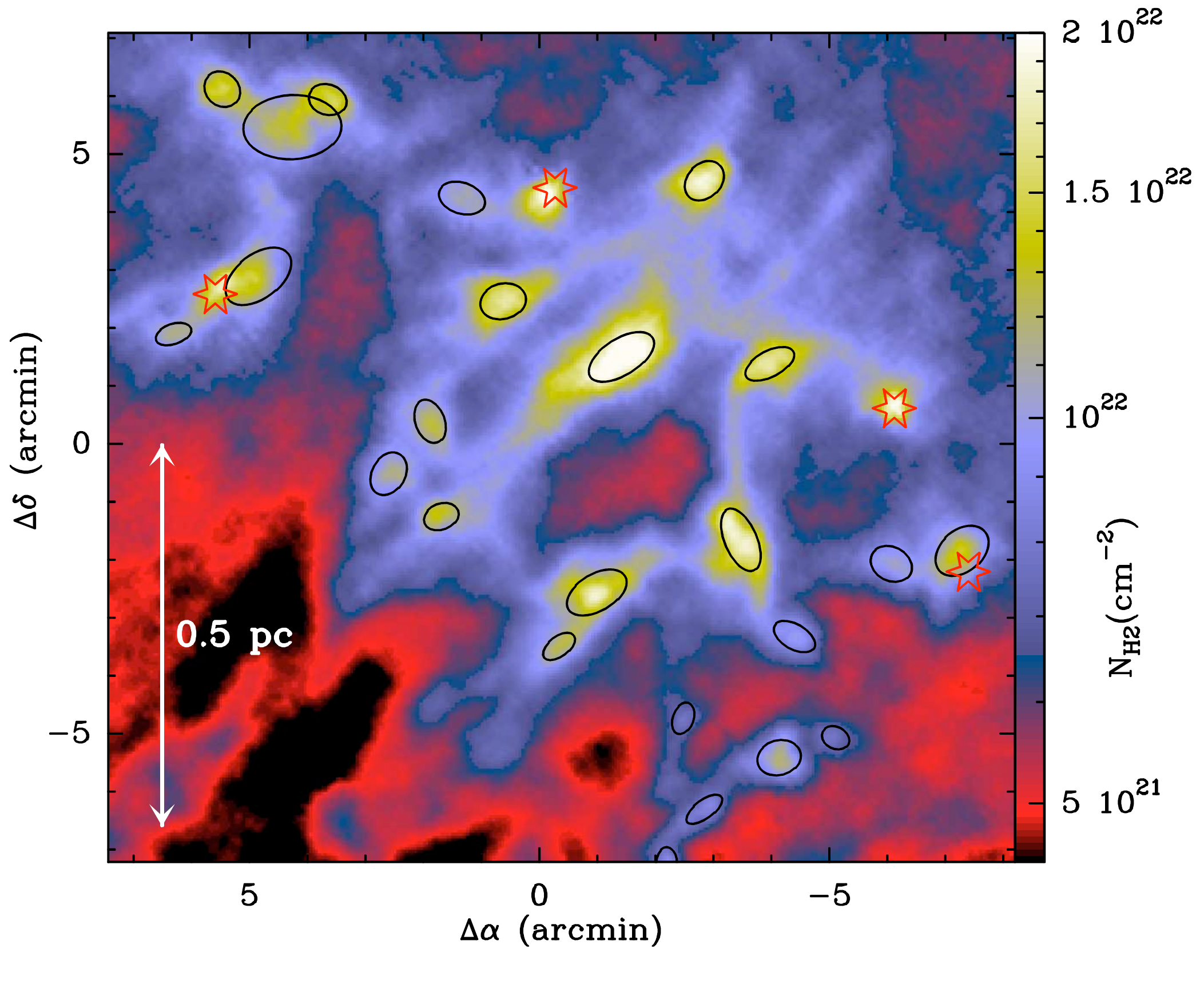} ~
\includegraphics[scale=0.475]{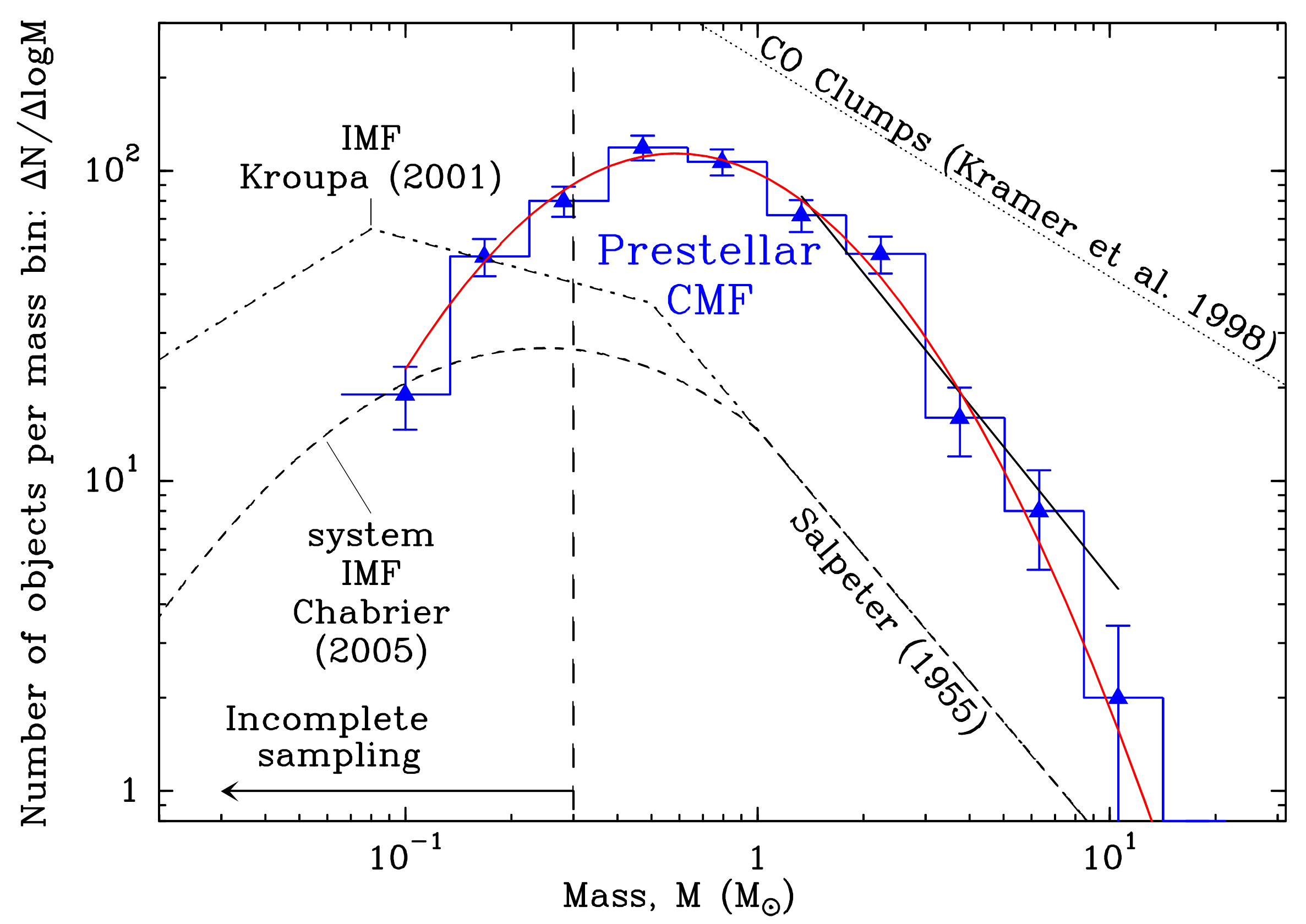}  
\caption{{\bf Top:} Close-up column density image of a small subfield in the Aquila Rift complex 
showing several candidate prestellar cores identified with $Herschel$ (adapted from K\"onyves et al. 2010).
The black ellipses mark the major and minor FWHM sizes determined for these cores 
by the source extraction algorithm {\it getsources} (Menshchikov et al. 2010, 2012).
Four protostellar cores are also shown by red stars. The effective  resolution of the  image 
is $\sim 18"$  or $\sim 0.02$~pc at $d \sim 260$~pc. 
{\bf Bottom:} Core mass function (blue histogram with error bars) of the $\sim 500$ candidate prestellar cores identified with $Herschel$ 
in Aquila (Andr\'e et al. 2010 and K\"onyves et al. 2010). 
The IMF of single stars (corrected for binaries -- e.g. Kroupa 2001),  the IMF of multiple systems (e.g. Chabrier 2005), 
and the typical mass spectrum of CO clumps (e.g. Kramer et al. 1998) are shown for comparison.
A log-normal fit to the observed CMF is superimposed (red curve); it peaks at $\sim 0.6\, M_\odot $, close to the 
Jeans mass within marginally critical filaments at $T \sim 10$~K (cf. \S ~6).
\label{aquila_cmf}
}
\end{figure}

Using  {\it getsources},  more than 200 starless cores but no protostars were detected in the $Herschel$ images 
of the Polaris flare region ($\sim 8$~deg$^2$ field). 
The locations of the Polaris starless cores in a mass versus size diagram show that 
they are $\sim 2$ orders of magnitude less dense than self-gravitating isothermal Bonnor-Ebert 
spheres and therefore cannot be gravitationally bound.  
The mass function of these unbound starless cores peaks at an order 
of magnitude smaller mass than the stellar IMF 
(Andr\'e et al. 2010). 
In contrast, more than 200 (Class 0 \& Class I) protostars could be identified in the $Herschel$ images of the whole ($\sim 11$~deg$^2$) 
Aquila region (Bontemps et al. 2010, Maury et al. 2011), along with more than 500 
starless cores $\sim$~0.01--0.1~pc in size (see Fig.~5 and Fig.~6a for some examples). 
Most ($> 60\% $) of the Aquila 
starless cores lie close to the loci of critical Bonnor-Ebert spheres 
in a mass versus size diagram, suggesting that they are self-gravitating and prestellar in nature 
(K\"onyves et al. 2010). 
The CMF derived for the entire sample of $> 500 $  
starless cores in Aquila is well fit by a log-normal distribution and 
closely resembles the IMF (Fig.~6b -- K\"onyves et al. 2010; Andr\'e et al. 2010). 
The similarity between the Aquila CMF and the Chabrier (2005) system IMF is consistent with an essentially one-to-one 
correspondence between core mass and stellar system mass ($M_{\star \rm sys} = \epsilon_{\rm core}\, M_{\rm core} $).
Comparing the peak of the CMF to the peak of the system IMF suggests that the efficiency $ \epsilon_{\rm core} $ 
of the conversion from core mass to stellar system mass is between $\sim 0.2$ and $\sim 0.4$ in Aquila. 

The first results of the HGBS 
survey on this topic therefore confirm the existence of a close relationship between the 
prestellar CMF and the stellar IMF, using data with already a factor of $\sim $ 2 to 9 better counting statistics 
than earlier ground-based studies (cf. Motte, Andr\'e, Neri 1998; Johnstone et al. 2000; Stanke et al. 2006; Alves et al. 2007; Enoch et al. 2008). 
The efficiency factor $ \epsilon_{\rm core} \sim 30\% $ may be attributed to mass loss due to the effect of outflows during the 
protostellar phase (Matzner \& McKee 2000). 
More work is needed to derive a reliable prestellar CMF at the low-mass end and fully assess the potential importance of subtle 
observational biases (e.g. background-dependent incompleteness and blending of unresolved groups of cores).
The results from the entire Gould Belt survey will also be necessary to fully characterize the nature of the CMF--IMF relationship 
as a function of environment.
Our early findings with $Herschel$ nevertheless seem to support models of the IMF based on pre-collapse cloud fragmentation 
such as the gravo-turbulent fragmentation picture (e.g. Larson 1985;  
Klessen \& Burkert 2000; Padoan \& Nordlund 2002; Hennebelle \& Chabrier 2008). 
Independently of any model, the $Herschel$ observations  suggest that one of the keys to the problem of the origin of the IMF  
lies in a good understanding of the formation mechanism of prestellar cores. 
This is true even if additional processes, such as rotational subfragmentation of prestellar 
cores into 
multiple systems during collapse (Bate et al. 2003; Goodwin et al. 2008) and 
``competitive'' accretion from a larger-scale mass reservoir 
at the protostellar stage (e.g. Bate \& Bonnell 2005), 
probably also play some role  and help to populate the low- and high-mass ends of the IMF, respectively. 
In Sect.~6 below, we argue that the prestellar cores responsible for  
the peak of the CMF/IMF result primarily from filament fragmentation.  

\section{Theoretical considerations on filament collapse and fragmentation}
The collapse and fragmentation properties of filaments under the assumption of cylindrical symmetry 
are well known theoretically  (e.g. Nagasawa 1987) but have received renewed attention with the $Herschel$ results. 
The gravitational instability of nearly isothermal filaments is primarily controlled by the value of their mass per unit length 
$M_{\rm line} \equiv M/L$. 
Above the critical value $M_{\rm line, crit} = 2\, c_s^2/G$ (where $c_s $ is the isothermal sound speed) cylindrical filaments are expected 
to be globally unstable to both radial collapse and fragmentation along their lengths (e.g. Inutsuka \& Miyama 1992, 1997), 
while below $M_{\rm line, crit} $ filaments are gravitationally unbound and thus expected to expand into the surrounding medium 
unless they are confined by some external pressure (e.g. Fischera \& Martin 2012). 
Note that the critical mass per unit length $M_{\rm line, crit}  \approx 16\, M_\odot /{\rm pc} \times (T_{\rm gas}/10\, {\rm K}) $
depends only on gas temperature $T_{\rm gas}$ (Ostriker 1964). 
In the presence of non-thermal gas motions, the critical mass per unit length becomes 
$M_{\rm line, vir} = 2\, \sigma_{\rm tot}^2/G $, 
also called the virial mass per unit length, where $\sigma_{\rm tot} = \sqrt{c_s^2 + \sigma_{\rm NT}^2} $ 
is the total one-dimensional velocity dispersion including both thermal and non-thermal components (Fiege \& Pudritz 2000).
Furthermore, Fiege \& Pudritz (2000) have shown that  $M_{\rm line, vir}$ is only slightly modified in the case of magnetized filaments:
$M_{\rm line, vir}^{\rm mag} = M_{\rm line, vir}^{\rm unmag} \times  \left(1 - \mathcal{M}/|\mathcal{W}| \right)^{-1}$, 
where $\mathcal{M} $ is the magnetic energy (positive for poloidal magnetic fields and negative for toroidal fields) 
and $\mathcal{W} $ is the gravitational energy. In practice, since molecular clouds typically have $|\mathcal{M}|/|\mathcal{W}| \simlt 1/2 $ 
(Crutcher 1999, 2012), 
$M_{\rm line, vir}^{\rm mag} $ differs from $M_{\rm line, vir}^{\rm unmag} \equiv  2\, \sigma_{\rm tot}^2/G $ by less than a factor of 2.

Importantly, filaments differ from both sheets and spheroids in their global gravitational instability properties. 
For a sheet-like cloud, there is always an equilibrium configuration since the internal pressure gradient can always become strong 
enough to halt the gravitational collapse of the sheet independently of the initial state (e.g. Miyama et al. 1987; Inutsuka \& Miyama 1997). 
In contrast, the radial collapse of an isothermal cylindrical cloud cannot be halted and no equilibrium is possible
when the line mass exceeds the critical mass per unit length $M_{\rm line, crit}$. Conversely, if the line mass of the filamentary cloud 
is less than $M_{\rm line, crit}$, gravity can never be made to dominate by increasing the external pressure, so that the collapse is 
always halted at some finite cylindrical radius. 
Filaments also differ markedly from isothermal spherical clouds which can always be induced to collapse by 
a sufficient increase in external pressure (e.g. Bonnor 1956; Shu 1977). 
The peculiar behavior of the filamentary geometry in isothermal collapse is due to 
the fact that the isothermal equation of state ($\gamma = 1$) is a critical case for the collapse of a filament (e.g. Larson 2005):  
For a polytropic equation of state ($P \propto \rho^\gamma$) with $\gamma < 1$, an unstable cylinder can collapse indefinitely 
toward its axis, while if $\gamma > 1$ the pressure gradient increases faster than gravity during contraction and the collapse 
is always halted at a finite radius. 
For comparison, the critical value is $\gamma = 0$ for sheets and $\gamma = 4/3$ for spheres. 
Indefinite, global gravitational collapse of a structure can occur when $\gamma$ is smaller than the critical value and 
is suppressed when $\gamma$ is larger than the critical value. 
Gravitational fragmentation thus tends to be favored over global collapse 
when $\gamma$ is close to or larger than the critical value. 
Since actual molecular clouds are well described by an effective equation of state with $\gamma \simlt 1$ (see, e.g., Fig.~3b), 
this led Larson (2005) to suggest that the filamentary geometry may play a key role in cloud fragmentation leading to star formation 
(see also Nakamura 1998).

The fragmentation properties of filaments and sheets differ from those of spheroidal clouds in that there is 
a preferred scale for gravitational fragmentation which directly scales with the scale height of the filamentary or sheet-like medium (e.g. Larson 1985). 
In the spherical case, the largest possible scale or mode (i.e., overall collapse of the medium) has the fastest growth rate 
so that global collapse tends to overwhelm the local collapse of finite-sized density perturbations,  
and fragmentation is generally suppressed in the absence of sufficiently large initial density enhancements (e.g. Tohline 1982). 
It also well known that spherical collapse quickly becomes strongly centrally concentrated (Larson 1969; Shu 1977), 
which tends to produce a single central density peak as opposed to several condensations (e.g. Whitworth et al. 1996). 
In contrast, sheets have a natural tendency to fragment into filaments (e.g. Miyama et al. 1987) and filaments with line masses 
close to $M_{\rm line, crit}$ have a natural tendency to fragment into spheroidal cores (e.g. Inutsuka \& Miyama 1997). 
The filamentary geometry is thus the most favorable configuration for small-scale perturbations to collapse locally and 
grow significantly before global collapse overwhelms them (Pon et al. 2011; Toal\'a et al. 2012). 
To summarize, theoretical considerations alone emphasize the potential importance of filaments for core and 
star formation.

\section{The key role of filaments in the star formation process}

The quasi ``universal'' filamentary structure of the cold ISM (cf. \S ~2) must be closely related to the star formation 
process since more than $ 70\% $ of the prestellar cores identified with $Herschel$  in nearby clouds appear 
to lie within filaments. 
The remarkable correspondence between the spatial distribution of compact cores and the most prominent filaments 
(see Fig.~5 and Men'shchikov et al. 2010) suggests that {\it prestellar dense cores form primarily along filaments}.
More precisely, the prestellar cores identified with $Herschel$ are preferentially found within the {\it densest filaments} 
with column densities exceeding $\sim 7 \times 10^{21}$~cm$^{-2}$ 
(Andr\'e et al. 2010 and Fig.~5). 
In the Aquila region, for instance, the distribution of background cloud column densities for the population of prestellar cores 
shows a steep rise above $N_{\rm H_2}^{\rm back} \sim 5 \times 10^{21}$~cm$^{-2}$ (cf. Fig.~7) and is such that 
$\sim 90\% $ of the candidate bound cores are found above a background column density 
$N_{\rm H_2}^{\rm back} \sim 7 \times 10^{21}$~cm$^{-2}$, corresponding to a background 
visual extinction $A_V^{\rm back} \sim 8$. 
The $Herschel$ observations of the Aquila Rift complex therefore strongly support the existence of a column density or visual extinction 
threshold for the formation of prestellar cores at $A_V^{\rm back} \sim $~5--10, which had been suggested based on earlier 
ground-based studies of, e.g., the Taurus and Ophiuchus clouds (cf. Onishi et al. 1998; Johnstone et al. 2004).
In the Polaris flare cirrus, the observations are also consistent with such an extinction threshold since the observed background column densities 
are all below $A_V^{\rm back} \sim 8$ and there are no examples of bound prestellar cores in this cloud.
More generally, the results obtained with $Herschel$ in nearby clouds suggest that 
a fraction $f_{\rm pre} \sim \,$15--20\% of the total gas mass 
above the column density threshold is in the form of prestellar cores.

\begin{figure}[!h]
\center
\hspace{-1mm}
  \resizebox{6.5cm}{!}{     
\includegraphics[angle=0,scale=0.7]{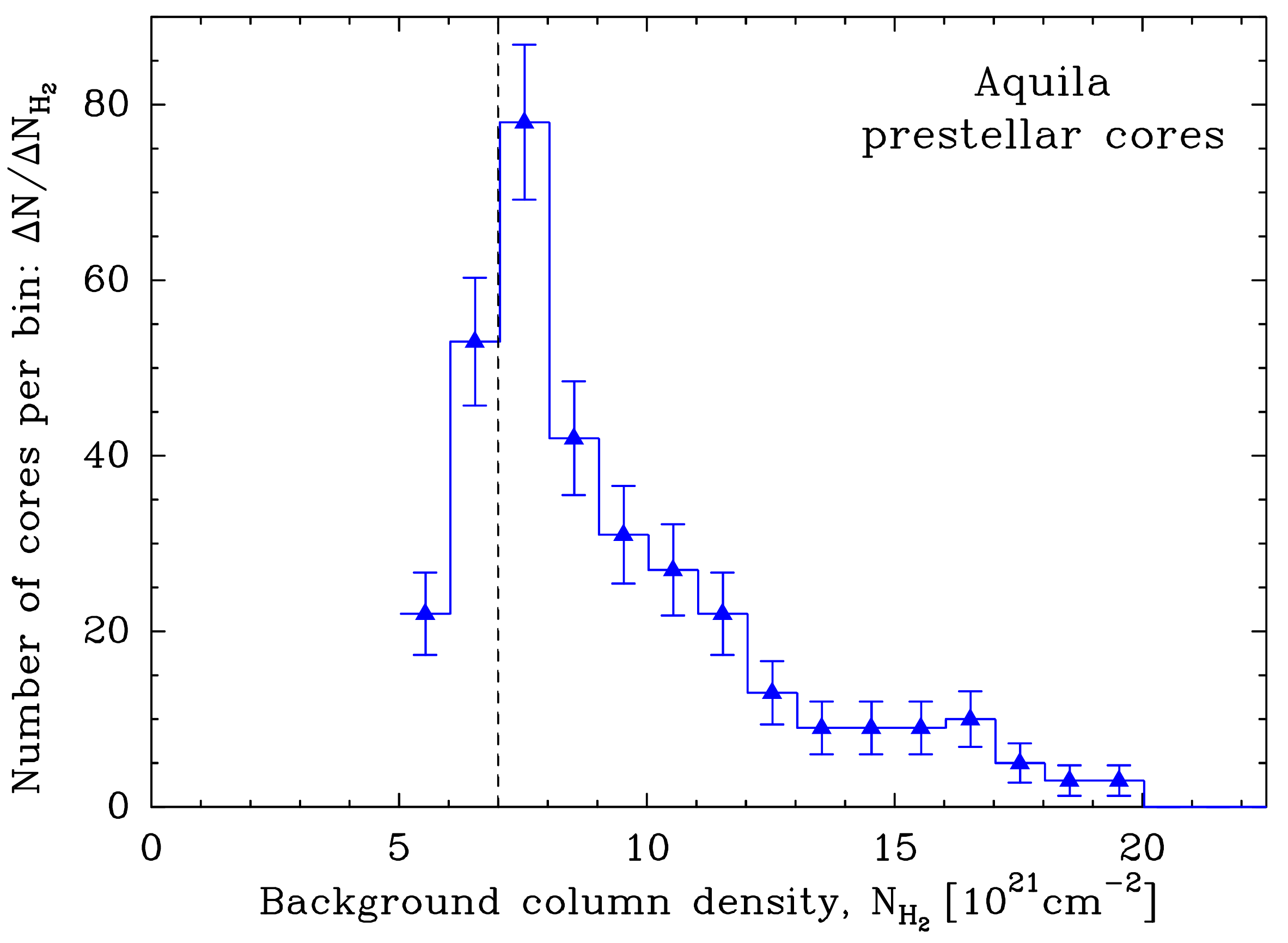}} 
 \resizebox{6.7cm}{!}{\includegraphics[angle=0,scale=0.7]{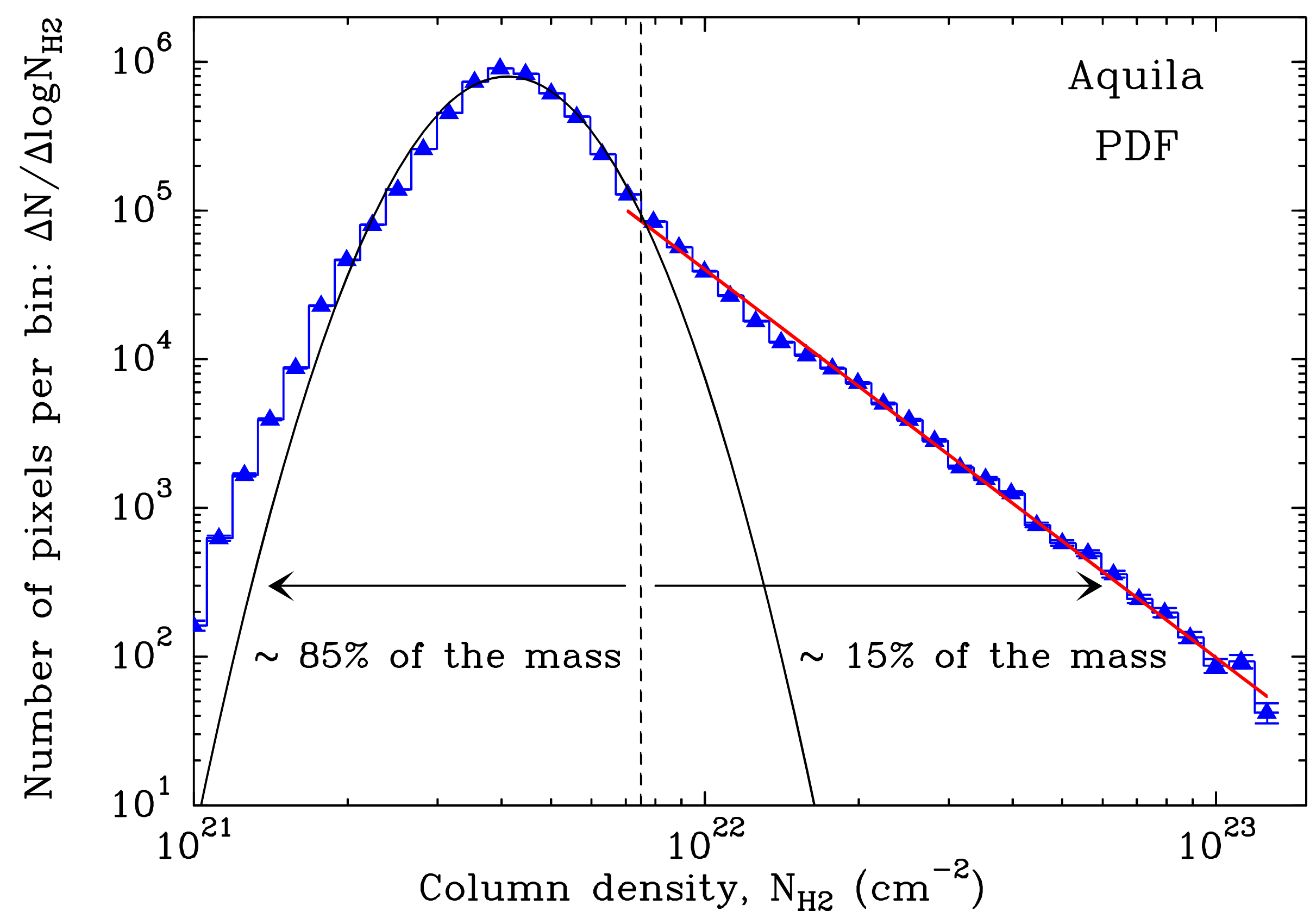}}
\caption{
{\bf Left:} Distribution of background column densities  for the 
candidate prestellar cores identified with $Herschel$ in the Aquila Rift complex (cf. K\"onyves et al. 2010).
The vertical dashed line marks the column density or extinction threshold 
at $N_{\rm H_2}^{\rm back} \sim 7 \times 10^{21}$~cm$^{-2}$ or $A_V^{\rm back} \sim 8$ 
(also corresponding to $\Sigma_{\rm gas}^{\rm back} \sim $~130~$M_\odot \, {\rm pc}^{-2} $). 
{\bf Right:} Probability density function of column density in the Aquila cloud complex, based on the column density image
derived from $Herschel$ data (Andr\'e et al. 2011; Schneider et al. 2013). A log-normal fit at low column densities and a power-law fit at high 
column densities are superimposed. The vertical dashed line marks the same column density threshold as in the left panel.
\label{background_ext}
}
\end{figure}

These $Herschel$ findings connect very well with the theoretical expectations for the gravitational instability of filaments (cf. \S~4) 
and point to an {\it explanation} of the star formation threshold in terms of the filamentary structure of molecular clouds. 
Given the typical width $W_{\rm fil} \sim 0.1$~pc measured for interstellar filaments (Arzoumanian et al. 2011; see Fig.~4) 
and the relation $M_{\rm line} \approx \Sigma_0 \times W_{\rm fil}$ between the central gas surface density $\Sigma_0$ 
and the line mass $M_{\rm line}$ of a filament, the threshold at $A_V^{\rm back} \sim 8$ or $\Sigma_{\rm gas}^{\rm back} \sim $~130~$M_\odot \, {\rm pc}^{-2} $ 
corresponds to within a factor of $< 2$ to the critical mass per unit length $M_{\rm line, crit} = 2\, c_s^2/G \sim 16\, M_\odot$/pc 
of nearly isothermal, long cylinders (see \S ~4 and Inutsuka \& Miyama 1997)  
for a typical gas temperature $T \sim 10$~K.
Thus, the core formation threshold approximately corresponds to the {\it threshold above which interstellar filaments are gravitationally unstable} 
(Andr\'e et al. 2010). 
Prestellar cores tend to be observed only above this threshold (cf. Figs.~5 \& 7) because they form out of a filamentary background and only the supercritical 
filaments with $ M_{\rm line} > M_{\rm line, crit} $ are able to fragment into self-gravitating cores.

For several reasons, the column density threshold for core and star formation within filaments is not a sharp boundary but a smooth transition.

\noindent
First, observations only provide information on the {\it projected} column density 
$\Sigma_{\rm obs} = \frac{1}{{\rm cos} \,i} \, \Sigma_{\rm int}$ of any given filament, where $i$ is the inclination angle to the plane of the sky 
and $\Sigma_{\rm int}$ is the intrinsic column density of the filament (measured perpendicular to its long axis).
For a population of randomly oriented filaments with respect to the plane of the sky, 
the net effect is that $\Sigma_{\rm obs} $ {\it overestimates} $\Sigma_{\rm int} $ by a factor  
$<\frac{1}{{\rm cos}\,i}>\, = \frac{\pi}{2}\,\sim1.57$ on average. Likewise, the apparent masses per unit length tend 
to slightly overestimate the intrinsic line masses of the filaments.  
However, the probability density function of the correction factor is such that the median correction is only  $\sim 15\% $ 
and that the correction is less than a factor of 1.5 for 75\% of the filaments. 
Although systematic, this projection effect thus remains small and has little impact on the global classification of observed 
filamentary structures as supercritical or subcritical filaments. 

\begin{figure}[!h]
\center
\includegraphics[scale=0.4]{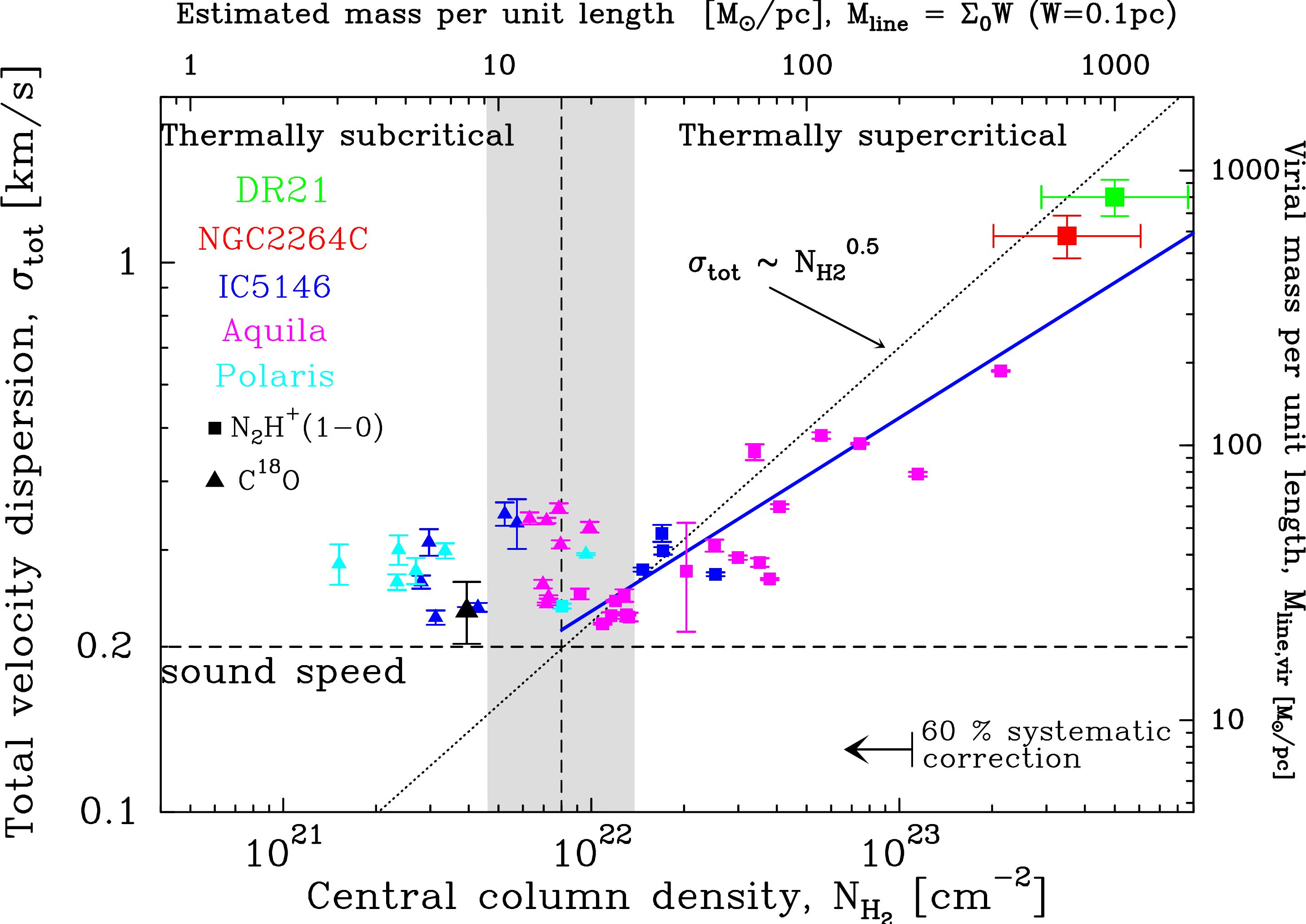}
\caption{
Total (thermal $+$ nonthermal) velocity dispersion versus central  column density for a sample of 
46 filaments in nearby interstellar clouds (from Arzoumanian et al. 2013).
The horizontal dashed line shows the value of the thermal sound speed $\sim 0.2$ km/s for  $T$~=~10~K. 
The vertical grey band marks the border zone between thermally subcritical and thermally supercritical filaments 
where the observed mass per unit length $ M_{\rm line} $ 
is close to the critical value $ M_{\rm line,crit} \sim$ 16~M$_{\odot}$/pc for $T$~=~10~K.
The blue  solid  line shows the best power-law fit $ \sigma_{\rm tot} \propto {N_{\rm H_2}}^{0.36~\pm~0.14}$
to the data points corresponding to  supercritical filaments. 
\label{veldisp_coldens}
}
\end{figure}

Second there is also a spread in the distribution of filament inner widths of about a factor of 2 on either side of 0.1~pc 
(Arzoumanian et al. 2011 -- cf. Fig. 4), implying a similar spread in the intrinsic column densities corresponding to the 
critical filament mass per unit length $M_{\rm line, crit} $. 

\noindent 
Third, interstellar filaments are not all exactly at $T = 10$~K and their internal velocity dispersion sometimes 
includes a small nonthermal component, $\sigma_{\rm NT}$, which must be accounted for in the evaluation 
of the critical or virial mass per unit length (Fiege \& Pudritz 2000): 
$M_{\rm line, vir} = 2\, \sigma_{\rm tot}^2/G $, where $\sigma_{\rm tot} $  
is the total internal velocity dispersion (see \S ~4).
Recent molecular line measurements with the IRAM 30m telescope for a sample of $Herschel$ filaments in 
the Aquila, IC5146, and Polaris clouds (Arzoumanian et al. 2013 -- cf. Fig.~8) show that both thermally subcritical filaments and nearly critical filaments 
(with $ M_{\rm line}$ within a factor of two of the thermal value of the critical mass per unit length $ M_{\rm line, crit} $) have ``transonic''  internal velocity dispersions 
$\sigma_{\rm tot}$ such that $c_s \simlt \sigma_{tot} < 2\, c_s $. 
Only the densest filaments (with $ M_{\rm line} >> M_{\rm line, crit} $) have internal velocity dispersions significantly in excess 
of the thermal sound speed  $c_s \sim 0.2$~km/s (cf. Fig.~8). These velocity dispersion measurements confirm that there is a critical threshold 
in mass per unit length above which interstellar filaments are self-gravitating and below which they are unbound, and that the position of this 
threshold lies between $\sim 16\, M_\odot$/pc and $\sim 32\, M_\odot$/pc, i.e., is consistent to within a factor of 2 with the thermal value of the 
critical mass per unit length $M_{\rm line,crit}$ for $T = 10$~K (Arzoumanian et al. 2013). 
The IRAM results shown in Fig.~8 thus confirm that the thermal critical mass per unit length $M_{\rm line,crit}$ 
plays a fundamental role in the evolution of interstellar filaments. Combined with the $Herschel$ findings summarized above 
and  illustrated in Fig.~5 and Fig.~7, this supports the view that the gravitational fragmentation of filaments may control the bulk 
of core and star formation, at least in nearby Galactic clouds.

\section{Implications for the IMF and the global star formation rate}

Since most stars appear to form in filaments, the fragmentation of filaments at the threshold of gravitational instability is a plausible 
mechanism for the origin of (part of) the stellar IMF. We may expect local collapse into spheroidal protostellar cores to be controlled by the 
Jeans/Bonnor-Ebert criterion $M \geq M_{\rm BE}$ where the Jeans or critical Bonnor-Ebert mass $M_{\rm BE}$ (e.g. Bonnor 1956)  
is given by: 
$$ M_{\rm BE} \sim 1.3\, c_s^4 /G^2 \Sigma_{\rm cl} \sim 0.53\, M_\odot\, \times \left(T_{\rm eff} /10\, {\rm K}\right)^2 \times  \left(\Sigma_{\rm cl}/160\, M_\odot\, {\rm pc}^{-2}\right)^{-1}.\\ (2)$$

\noindent
If we consider a quasi-equilibrium isothermal cylindrical filament on the verge of {\it global} radial collapse, 
it has a mass per unit length equal to the critical value $M_{\rm line, crit} = 2\, c_s^2/G$ ($\sim 16\, M_\odot$/pc for $T_{\rm eff} \sim 10$~K) and an effective 
diameter $D_{\rm flat, crit} = 2\, c_s^2/G\Sigma_0$ ($\sim 0.1\, $pc for $T_{\rm eff} \sim 10$~K and $\Sigma_0 \sim 160\, M_\odot\, {\rm pc}^{-2}$).
A segment of such a cylinder of length equal to $D_{\rm flat, crit}$ contains a mass 
$M_{\rm line, crit} \times D_{\rm flat, crit} = 4\, c_s^4/G^2\Sigma_0 \sim 3 \times M_{\rm BE} $ 
($\sim 1.6\, M_\odot$ for $T_{\rm eff} \sim 10$~K and $\Sigma_0 \sim 160\, M_\odot\, {\rm pc}^{-2}$) and is thus potentially prone to 
{\it local} Jeans instability. 
Since local collapse tends to be favored over global collapse in the case of nearly isothermal filaments 
(see Sect.~4 and Pon et al. 2011), gravitational fragmentation 
into spheroidal cores is expected to occur along supercritical filamentary structures, 
as indeed found in both numerical simulations (e.g. Bastien et al. 1991; Inutsuka \& Miyama 1997) 
and $Herschel$ observations (see Fig.~5 and Sect.~5).  
Remarkably, 
the peak of the prestellar CMF at $ \sim 0.6\, M_\odot $ as observed in the Aquila cloud complex (cf. Fig.~6b) 
corresponds very well to the Bonnor-Ebert mass $ M_{\rm BE} \sim 0.5\, M_\odot $ 
within marginally critical filaments with $ M_{\rm line} \approx M_{\rm line, crit} \sim 16\, M_\odot$/pc and surface densities 
$\Sigma \approx \Sigma_{\rm gas}^{\rm crit} \sim 160\, M_\odot \, {\rm pc}^{-2} $. 
Likewise,  the median projected spacing $\sim 0.08$~pc observed between the prestellar cores of Aquila roughly matches the 
thermal Jeans length within marginally critical filaments.
All of this is consistent with the idea that gravitational fragmentation is the dominant physical mechanism generating prestellar cores 
within interstellar filaments. Furthermore, a typical prestellar core mass of $\sim 0.6\, M_\odot$  translates into a characteristic star 
or stellar system mass of $\sim 0.2\, M_\odot$, assuming a typical efficiency $ \epsilon_{\rm core} \sim 30\%$ (cf. \S ~3). 

Therefore, the $Herschel$ results strongly support Larson (1985)'s interpretation of the peak of the IMF in terms 
of the typical Jeans mass in star-forming clouds. 
Overall, our $Herschel$ findings 
suggest that the gravitational fragmentation of supercritical filaments produces the peak of the prestellar CMF which, 
in turn, may account for the log-normal ``base'' (cf. Bastian et al. 2010) of the IMF. 
It remains to be seen, however, whether the bottom end of the IMF and the Salpeter power-law slope at the high-mass end can also 
be explained by filament fragmentation. 
Naively, one would indeed expect gravitational fragmentation to result in a narrow prestellar CMF, sharply peaked at the median thermal
Jeans mass. It should also be noted that a small ($< 30\% $) fraction of prestellar cores do not appear to form along filaments. 
A good example is the pre-brown dwarf core Oph B-11 recently identified in the Ophiuchus main cloud 
(Greaves et al. 2003, Andr\'e et al. 2012). 
On the other hand, a Salpeter power-law tail at high masses may be produced by filament 
fragmentation if turbulence has generated an appropriate field of initial density fluctuations within the filaments in the first place (cf. Inutsuka 2001). 
More precisely, Inutsuka (2001) has shown that if the power spectrum of initial density fluctuations along the filaments approaches 
$P(k) \equiv |\delta_k|^2 \propto k^{-1.5} $ then the CMF produced by gravitational fragmentation evolves toward a power-law 
$dN/dM \propto M^{-2.5}$,  similar to the Salpeter IMF  ($dN/dM_\star  \propto M_\star^{-2.35} $).
Interestingly, the power spectrum of column density fluctuations along the filaments observed with $Herschel$ in nearby clouds 
is typically $P(k) \propto k^{-1.8} $, which is close to the required spectrum. 
Alternatively, a CMF with a Salpeter power-law tail may result from the gravitational fragmentation of a population of filaments
with a distribution of supercritical masses per unit length. 
Observationally, the supercritical filaments observed as part of the $Herschel$ Gould Belt survey do seem to have a power-law
distribution of masses per unit length $dN/dM_{\rm line}  \propto M_{\rm line}^{-2.2}$ above $\sim 20\, M_\odot$/pc. 
Since the width of the filaments is roughly constant ($W_{\rm fil}  \sim 0.1$~pc), the mass per unit length is directly proportional to 
the central surface density, $M_{\rm line} \sim \Sigma \times W_{\rm fil} $. 
Furthermore, the total velocity dispersion of these filaments increases roughly as $ \sigma_{\rm tot} \propto \Sigma^{0.5}$ (Arzoumanian et al. 2013 -- see Fig.~8), 
which means that their effective temperature scales roughly as $T_{\rm eff} \propto \Sigma $. 
Hence $M_{\rm BE} \propto  \Sigma \propto M_{\rm line} $, and the observed distribution of masses per unit length directly translates into 
a power-law distribution of Bonnor-Ebert masses $dN/dM_{\rm BE}  \propto M_{\rm BE}^{-2.2}$ along supercritical filaments, which is also reminiscent 
of the Salpeter IMF.
Although these two alternative possibilities seem promising, more work is needed to assess whether the direct gravitational fragmentation of filaments 
can account for the high-mass end of the CMF/IMF. 

The realization that prestellar core formation occurs primarily along gravitationally unstable filaments of roughly constant width $W_{\rm fil} \sim 0.1$~pc 
also has potential implications for our understanding of star formation on global Galactic and extragalactic scales. 
Remarkably, the critical line mass of a filament, $M_{\rm line, crit} = 2\, c_s^2/G$, depends only on gas temperature 
(i.e., $T \sim 10$~K for the bulk of molecular clouds, away from the immediate vicinity of massive stars) and is modified by 
only a factor of order unity for filaments with realistic levels of magnetization (Fiege \& Pudritz 2000 -- see Sect.~4). 
This may set a quasi-universal threshold for star formation in the cold ISM of galaxies at $M_{\rm line, crit} \sim 16\, M_\odot$/pc in terms of 
filament mass per unit length, or $M_{\rm line, crit}/W_{\rm fil} \sim 160\, M_\odot \, {\rm pc}^{-2} $ in terms of gas surface density, 
or $M_{\rm line, crit}/W_{\rm fil}^2 \sim 1600\, M_\odot \, {\rm pc}^{-3} $ in terms of gas density (the latter corresponding to a volume number density 
$n_{H_2} \sim 2 \times 10^4\, {\rm cm}^{-3} $). 
Indeed, recent near-/mid-infrared studies of the star formation rate as a function of gas surface density in both Galactic and extragalactic cloud complexes 
(e.g. Heiderman et al. 2010; Lada et al. 2010) show that the star formation rate tends to be linearly proportional to the 
mass of dense gas above a surface density threshold $\Sigma_{\rm gas}^{\rm th} \sim $~120--130~$M_\odot \, {\rm pc}^{-2} $ 
and drops to negligible values below $\Sigma_{\rm gas}^{\rm th} $ (see Gao \& Solomon 2004 for external galaxies). 
Note that this is essentially the {\it same} threshold as found with $Herschel$ for the formation of prestellar cores in nearby clouds (cf. \S ~5 and Figs.~7 \& 8).
Moreover, the relation between the star formation rate $SFR$ and the mass of dense gas $M_{\rm dense}$ above the threshold 
is estimated to be ${SFR} = 4.6 \times 10^{-8}\,  M_\odot \, {\rm yr}^{-1}\, \times \left(M_{\rm dense}/M_\odot \right) $ in nearby clouds (Lada et al. 2010), 
which is close to the relation ${SFR} = 2 \times 10^{-8}\,  M_\odot \, {\rm yr}^{-1}\, \times \left(M_{\rm dense}/M_\odot \right) $ 
found by Gao \& Solomon (2004) for galaxies. 
Both of these values are very similar to the star formation rate per unit mass of dense gas 
${SFR}/M_{\rm dense} = f_{\rm pre} \times \epsilon_{\rm core}/t_{\rm pre} \sim 0.15 \times 0.3/10^6 \sim 4.5 \times 10^{-8}\, M_\odot \, {\rm yr}^{-1}$ 
that we may derive based on the $Herschel$ results in the Aquila complex, 
by considering that only a fraction $f_{\rm pre} \sim 15\%$ of the gas mass above the column 
density threshold is in the form of prestellar cores (cf. \S ~5), that the local star formation efficiency at the level of an individual core is $ \epsilon_{\rm core} \sim 30\%$ (cf. \S ~3), 
and that the typical lifetime of the Aquila cores is $t_{\rm pre} \sim 10^6$~yr (K\"onyves et al., in prep.). 
Despite relatively large uncertainties, the agreement with the extragalactic value of Gao \& Solomon (2004) is surprisingly good, 
implying that there may well be a quasi-universal ``star formation law'' converting dense gas into stars above the threshold (see also Lada et al. 2012).

\section{Conclusions: Toward a universal scenario for star formation? }

The results obtained with $Herschel$ on nearby clouds and  summarized in the previous sections
provide key insight into the first phases of the star formation process. 
They emphasize the role of filaments and 
support  a scenario according to which the formation of prestellar cores occurs in two main steps (see, e.g., Andr\'e et al. 2010).
First, large-scale magneto-hydrodynamic (MHD) turbulence 
generates a whole network of filaments in the ISM (cf. Padoan et al. 2001);  second, the densest filaments fragment into 
prestellar cores by gravitational instability (cf. Inutsuka \& Miyama 1997) above a critical (column) density 
threshold corresponding to $ \Sigma_{\rm gas}^{\rm crit} \sim 150\, M_\odot \, {\rm pc}^{-2} $ ($A_V^{\rm crit} \sim 8$) 
or $ n_{\rm H_2}^{\rm crit} \sim 2 \times 10^4\,  {\rm cm}^{-3} $.

That the formation of filaments in the diffuse ISM represents the first step toward core/star formation is suggested by the filaments 
{\it already} being omnipresent in a gravitationally unbound, non-star-forming cloud such as Polaris (cf. Fig.~1, Men'shchikov et al. 2010, and 
Miville-Desch\^enes et al. 2010). This indicates that interstellar filaments are not produced by large-scale gravity and that 
their formation must precede star formation. It is also consistent with the view that the filamentary structure results primarily from the dissipation of  
large-scale MHD turbulence (cf. Padoan et al. 2001; Hily-Blant \& Falgarone 2007, 2009). 
In the picture proposed by Padoan et al. (2001), the dissipation of turbulence occurs in shocks and interstellar filaments correspond to dense, 
post-shock stagnation gas associated with compressed regions between interacting supersonic flows. 
One merit of this picture is that it can qualitatively account for the characteristic 
$\sim 0.1$~pc width of the filaments as measured with $Herschel$ (cf. Fig.~4):  the typical thickness of shock-compressed structures 
resulting from supersonic turbulence in the ISM is expected to be roughly the sonic scale of the turbulence which is $\sim 0.1$~pc in diffuse interstellar gas 
(cf. Larson 1981, Falgarone et al. 2009, and discussion in Arzoumanian et al. 2011). 
Direct evidence of the role of large-scale compressive flows has been found with $Herschel$ in the Pipe Nebula in the form of filamentary structures with 
asymmetric column density profiles which most likely result from compression by the winds of the Sco OB2 association (Peretto et al. 2012). 
However, turbulent compression alone would tend to form layers and is unlikely to directly produce filaments. 
A more complete picture, recently proposed by Hennebelle (2013), is that interstellar filaments result from a combination of 
turbulent shear and compression. Hennebelle (2013) further suggests that the filament width $\sim 0.1$~pc may correspond  
to the dissipation length of MHD waves due to ion-neutral friction. 

The second step appears to be the gravitational fragmentation of the densest filaments with supercritical masses 
per unit length ($ M_{\rm line} \ge M_{\rm line, crit} $) into self-gravitating prestellar cores (cf. \S ~5). 
In active star-forming regions such as the Aquila complex, most of the prestellar cores identified 
with $Herschel$ are indeed concentrated within supercritical filaments (cf. Fig.~5). In contrast, in non-star-forming clouds such as Polaris, 
all of the filaments have subcritical masses per unit length and only unbound starless cores are observed 
but no prestellar cores nor protostars (cf. Fig.~1). 
$Herschel$ observations indicate that even star-forming, supercritical filaments maintain roughly constant inner widths $\sim 0.1$~pc
while evolving (Arzoumanian et al. 2011 -- see Figs.~3 \& 4). At first sight, this seems surprising since supercritical filaments 
are unstable to radial collapse and are thus expected to undergo rapid radial contraction with time (e.g. Kawachi \& Hanawa 1998 -- see Sect.~4). 
The most likely solution to this paradox is that supercritical filaments are {\it accreting} additional background material while contracting. 
The increase in velocity dispersion with central column density observed for supercritical filaments (Arzoumanian et al. 2013 -- see Fig.~8)
is indeed suggestive of an increase in (virial) mass per unit length with time.   
More direct evidence of this accretion process for supercritical filaments exists in 
several cases in the form of low-density striations or sub-filaments observed perpendicular to the main filaments 
and apparently feeding them from the side. Examples include the B211/B213 filament in Taurus (Palmeirim et al. 2013 -- see Fig.~ 2), 
the Musca filament (Cox et al. 2013, in preparation), and the DR21 ridge in Cygnus~X (Schneider et al. 2010; Hennemann et al. 2012).
In the case of the Taurus filament (Fig.~2), the estimated accretion rate is such that it would take $\sim \, $1--2~Myr for the central 
filament to double its mass (Palmeirim et al. 2013). 
This accretion process supplies gravitational energy to supercritical filaments which is then converted into turbulent kinetic energy 
(cf. Heitsch et al. 2009 and 
Klessen \& Hennebelle 2010) and may explain the observed increase in velocity dispersion 
($ \sigma_{\rm tot} \propto {\Sigma_0}^{0.5}$ -- cf. Fig.~8). 
The central diameter of such accreting filaments is expected to be of order the effective Jeans length 
$D_{\rm J,eff} \sim 2\, \sigma_{\rm tot}^2/G\Sigma_0 $, which Arzoumanian et al. (2013) have shown to remain close to $\sim 0.1$~pc. 
Hence, through accretion of parent cloud material, supercritical filaments may keep roughly constant inner widths and remain 
in rough virial balance while contracting. This process may effectively prevent the global (radial) collapse of supercritical filaments 
and thus favor their fragmentation into cores (e.g. Larson 2005), in agreement with the $Herschel$ results (see Fig.~5).

The above scenario can explain the peak for the prestellar CMF and may account for the base of the stellar IMF (see Sect.~6 and Fig.~6b). 
It partly accounts for the general inefficiency of the star formation process since, even in active star-forming complexes such as Aquila 
(Fig.~5), only a small fraction of the total gas mass ($\sim 15\%$ in the case of Aquila -- see Fig.~7b) is above of the column density threshold, 
and only a small fraction $f_{\rm pre} \sim 15\%$ of the dense gas above the threshold is in the form of prestellar cores (see Sect.~5).
Therefore, the vast majority of the gas in a GMC ($\sim 98\% $ in the case of Aquila) does not participate in star formation at any given time 
(see also Heiderman et al. 2010 and Evans 2011). 
Furthermore, the fact that essentially the same ``star formation law'' is observed above the column density threshold in both Galactic clouds 
and external galaxies (see Sect.~6 and Lada et al. 2012) suggests that the star formation scenario sketched above is quasi-universal and 
may well apply to the ISM of all galaxies. 

To conclude, the $Herschel$ results discussed in this paper are extremely encouraging as they 
point to a unified picture of star formation on GMC scales in both Galactic clouds and external galaxies. 
Confirming and refining the scenario proposed here will require follow-up observations to constrain the dynamics of the filamentary structures
imaged with $Herschel$ as well as detailed comparisons with numerical simulations of molecular cloud formation and evolution. 
ALMA will be instrumental in testing whether this scenario, based on $Herschel$ observations of nearby Galactic clouds 
forming mostly low-mass stars, 
is truly universal and 
also applies to high-mass star forming clouds and 
the GMCs of other galaxies.

\begin{acknowledgments}
It is a pleasure for me to acknowledge the important contributions of my colleagues of the $Herschel$/SPIRE SAG~3 working group, e.g., Doris Arzoumanian, James Di Francesco, 
Jason Kirk, Vera K\"onyves, Sasha Men'shchikov, Fred Motte, 
Pedro Palmeirim, Nicolas Peretto, Nicola Schneider, Derek Ward-Thompson, to the results presented in this paper.
I am also grateful to Shu-ichiro Inutsuka, Fumitaka Nakamura, Patrick Hennebelle, Ralph Pudritz, Zhi-Yun Li, and Shantanu Basu for insightful discussions on filaments. 
This work has received support from the European Research Council under the European Union's Seventh 
Framework Programme (Grant Agreement no. 291294) and from the French National Research Agency (Grant no. ANR--11--BS56--0010).
\end{acknowledgments}


\begin{thebibliography}{}

\bibitem[Abergel et al.(1994)]{Abergel94} Abergel, A., Boulanger, F., Mizuno, A., \& Fukui, Y. 1994, ApJ, 423, L59
\bibitem[Alves et al.(2001)]{Alves01} Alves, J. F., Lada, C. J., \& Lada, E. A. 2001, Nature, 409, 159
\bibitem[Alves et al.(2007)]{Alves07} Alves, J. F., Lombardi, M., \& Lada, C. J. 2007, A\&A, 462, L17
\bibitem[Andr{\' e} et al.(2007)]{andre07} Andr\'e, P., Belloche, A., Motte, F., \& Peretto, N. 2007, A\&A, 472, 519
\bibitem[Andr{\' e} et al.(2010)]{Andre10} Andr\'e, Ph., Men'shchikov, A., Bontemps, S. et al. 2010, A\&A, 518, L102
\bibitem[Andr{\' e} et al.(2011)]{Andre11} Andr\'e, Ph., Men'shchikov, A., K\"onyves, V., \& Arzoumanian, D. 2011, in Computational Star Formation, IAU Symp. 270, Eds. J. Alves et al., p. 255
\bibitem[Andr{\' e} et al.(2000)]{Andre00} Andr{\' e}, P., Ward-Thompson,D., \& Barsony, M.\ 2000, in Protostars and Planets IV, Eds V. Mannings et al., p.59
\bibitem[Andr{\' e} et al.(2012)]{Andre12} Andr{\' e}, P., Ward-Thompson,D., \& Greaves, J.S.\ 2012, Science, 337, 69 
\bibitem[Arzoumanian et al.(2011)]{Arzoumanian11} Arzoumanian, D., Andr\'e, Ph., Didelon, P. et al. 2011, A\&A, 529, L6
\bibitem[Arzoumanian et al.(2013)]{Arzoumanian13} Arzoumanian, D., Andr\'e, Ph., Peretto, N., \&  K\"onyves, V. 2013, A\&A, 553, A119
\bibitem[Bastian et al.(2010)]{Bastian10} Bastian, N., Covey, K.R., \& Meyer, M.R. 2010, ARA\&A, 48, 339
\bibitem[Bastien et al.(1991)]{Bastien91} Bastien, P., Arcoragi, J.-P., Benz, W., Bonnell, I., \& Martel, H. 1991, ApJ, 378, 255
\bibitem[Bate \& Bonnell(2005)]{bate05}Bate, M. R., \& Bonnell, I. A. 2005, MNRAS, 356, 1201
\bibitem[Bate et al.(2003)]{Bate03} Bate, M. R., Bonnell, I. A., \& Bromm, V. 2003, MNRAS, 339, 577
\bibitem[Bergin \& Tafalla(2007)]{Bergin07} Bergin, E.A., \& Tafalla, M. 2007, ARA\&A, 45, 339
\bibitem[Bonnor(1956)]{Bonnor56} Bonnor, W.B. 1956, MNRAS, 116, 351
\bibitem[Bontemps et al.(2010)]{Bontemps10} Bontemps, S., Andr\'e, Ph., K\"onyves, V. et al. 2010, A\&A, 518, L85
\bibitem[Chabrier(2005)]{chabrier05} Chabrier, G. 2005, in The Initial Mass Function 50 years later, Eds. E. Corbelli et al., p.41
\bibitem[Chapman et al.(2011)]{chapman11} Chapman, N.L., Goldsmith, P.F., Pineda, J.L. et al. 2011, ApJ, 741, 21
\bibitem[Crutcher(1999)]{crutcher99} Crutcher, R. M. 1999, ApJ, 520, 706  
\bibitem[Crutcher(2012)]{crutcher12} Crutcher, R. M. 2012, ARA\&A, 50, 29
\bibitem[Di Francesco et al.(2007)]{difrancesco07} Di Francesco, J., Evans II, N.J., Caselli, P. et al.
2007, in Protostars and Planets V, p. 17
\bibitem[Dunham et al.(2008)]{Dunham08} Dunham, M.M., Crapsi, A., Evans, N.J. et al.  2008, ApJS, 179, 249
\bibitem[Enoch et al.(2008)]{Enoch08} Enoch, M. L., Young, K. E., Glenn, J., Evans, N. J. et al.  2008, ApJ, 684, 1240
\bibitem[Evans(2011)]{Evans11} Evans, N.J.  2011, in Computational Star Formation, IAU Symp. 270, Eds. J. Alves et al., p. 25
\bibitem[Falgarone et al.(2009)]{Falgarone09} Falgarone, E., Pety, J., \& Hily-Blant, P. 2009, A\&A, 507, 355
\bibitem[Federrath et al.(2010)]{Federrath10} Federrath, C., Roman-Duval, J., {Klessen}, R.S. et al. 
2010, A\&A, 512, A81 
\bibitem[Fiege \& Pudritz(2000)]{Fiege00} Fiege, J.D., \& Pudritz, R.E. 2000,  MNRAS, 311, 85
\bibitem[Fischera \& Martin(2012)]{Fischera12} Fischera, J., \& Martin, P.G. 2012, A\&A, 542, A77
\bibitem[Gao \& Solomon(2004)]{Gao04} Gao, Y., \& Solomon, P. 2004, ApJ, 606, 271
\bibitem[Gong \& Ostriker(2011)]{Gong11} Gong, H., \& Ostriker, E.C. 2011, ApJ, 729, 120
\bibitem[Goodman et al.(1998)]{Goodman98} Goodman, A. A., Barranco, J. A., Wilner, D. J., \& Heyer, M. H. 1998, ApJ, 504, 223
\bibitem[Goodwin et al.(2008)]{Goodwin08} Goodwin, S. P., Nutter, D., Kroupa, P., Ward-Thompson, D., Whitworth, A. P. 2008, A\&A, 477, 823
\bibitem[Greaves et al.(2003)]{Greaves03} Greaves, J.S., Holland, W.S., \& Pound, M.W. 2003, MNRAS, 346, 441
\bibitem[Guillout(2001)]{Guillout01} Guillout, P. 2001, in From Darkness to Light, Eds. T. Montmerle \& P. Andr\'e,
ASP Conf. Ser., 243, p. 677
\bibitem[Gutermuth et al.(2008)]{Gutermuth08} Gutermuth, R.A., Bourke, T.L., Allen, L.E. et al. 2008, ApJ, 673, L151
\bibitem[Hacar \& Tafalla(2011)]{Hacar2011} Hacar, A., \& Tafalla, M. 2011, A\&A, 533, A34
\bibitem[Hartmann(2002)]{Hartmann02} Hartmann, L. 2002, ApJ, 578, 914
\bibitem[Hatchell et al.(2005)]{hatchell05} Hatchell, J.,  Richer, J. S.,  Fuller, G. A. et al. 2005, A\&A, 440, 151
\bibitem[Heiderman et al.(2010)]{Heiderman10} Heiderman, A., Evans, N.J., Allen, L.E. et al. 2010, ApJ, 723, 1019
\bibitem[Heithausen et al.(2002)]{Heithausen02} Heithausen, A. et al. 2002, A\&A, 383, 591
\bibitem[Heitsch et al.(2009)]{Heitsch09} Heitsch, F., Ballesteros-Paredes, J., \& Hartmann, L. 2009, ApJ, 704, 1735
\bibitem[Hennebelle(2013)]{Hennebelle13} Hennebelle, P. 2013, A\&A, 556, A53
\bibitem[Hennebelle \& Chabrier(2008)]{Hennebelle08} Hennebelle, P., \& Chabrier, G. 2008, ApJ, 684, 395
\bibitem[Hennemann et al.(2012)]{Hennemann12} Hennemann, M., Motte, F., Schneider, N. et al. 2012, A\&A, 543, L3
\bibitem[Heyer et al.(2008)]{Heyer08} Heyer, M., Gong, H., Ostriker, E. \& Brunt, C. 2008, ApJ, 680, 420
\bibitem[Hily-Blant \& Falgarone(2007)]{HilyBlant07} Hily-Blant, P., \& Falgarone, E. 2007, A\&A, 469, 173
\bibitem[Hily-Blant \& Falgarone(2009)]{HilyBlant09} Hily-Blant, P., \& Falgarone, E. 2009, A\&A, 500, L29
\bibitem[Inutsuka(2001)]{Inutsuka01} Inutsuka, S. 2001, {ApJ}, 559, L149
\bibitem[Inutsuka \& Miyama(1992)]{Inutsuka92} Inutsuka, S., \& Miyama, S.M. 1992, ApJ, 388, 392
\bibitem[Inutsuka \& Miyama(1997)]{Inutsuka97} Inutsuka, S., \& Miyama, S.M. 1997, ApJ, 480, 681
\bibitem[Johnstone et al.(2000)]{Johnstone00} Johnstone, D., Wilson, C.D., Moriarty-Schieven, G. et al. 2000, ApJ, 545, 327
\bibitem[Johnstone et al.(2004)]{Johnstone04} Johnstone, D., Di Francesco, J., \& Kirk, H. 2004, ApJ, 611, L45
\bibitem[Kawachi \& Hanawa(1998)]{Kawachi98} Kawachi, T., \& Hanawa, T. 1998, PASJ, 50, 577
\bibitem[Kennicutt(1998)]{Kennicutt98} Kennicutt, R. 1998, ApJ, 498, 541
\bibitem[Klessen \& Burkert(2000)]{Klessen00} Klessen, R. S., \& Burkert, A. 2000, ApJS, 128, 287
\bibitem[Klessen \& Hennebelle(2010)]{Klessen10} Klessen, R. S., \& Hennebelle, P. 2010, A\&A, 520, A17
\bibitem[K\"onyves et al.(2010)]{Konyves10} K\"onyves, V., Andr\'e, Ph., Men'shchikov, A. et al. 2010, A\&A, 518, L106
\bibitem[Kramer et al.(1998)]{Kramer98} Kramer, C., Stutzki, J., Rohrig, R., Corneliussen, U. 1998, A\&A, 329, 249
\bibitem[Kroupa(2001)]{kroupa01} Kroupa, P. 2001, MNRAS, 322, 231
\bibitem[Lada et al.(2010)]{Lada10} Lada, C.J., Lombardi, M, \& Alves, J. 2010, {ApJ}, 724, 687
\bibitem[Lada et al.(2012)]{Lada12} Lada, C.J., Forbrich, J., Lombardi, M, \& Alves, J. F. 2012, {ApJ}, 745, 190
\bibitem[Larson(1969)]{larson69} Larson, R. B. 1969, MNRAS, 145, 271
\bibitem[Larson(1981)]{Larson81} Larson, R.B., 1981, {MNRAS}, 194, 809
\bibitem[Larson(1985)]{Larson85} Larson, R.B. 1985, MNRAS, 214, 379
\bibitem[Larson(2005)]{Larson05} Larson, R. B. 2005, MNRAS, 359, 211
\bibitem[Li \& Goldsmith(2012)]{Li2012} Li, D., \& Goldsmith, P.F. 2012, ApJ, 756, 12
\bibitem[Matzner \& McKee(2000)]{Matzner00} Matzner, C.D., \& McKee, C.F. 2000, ApJ, 545, 364
\bibitem[Maury et al.(2011)]{Maury11} Maury, A., Andr\'e, Ph., Men'shchikov, A., K\"onyves, V., \& Bontemps, S.  2011, A\&A, 535, A77
\bibitem[Men'shchikov et al.(2010)]{Menshch10} Men'shchikov, A., Andr\'e, Ph., Didelon, P. et al. 2010, A\&A, 518, L103
\bibitem[Men'shchikov et al.(2012)]{Menshch12} Men'shchikov, A. Andr\'e, Ph., Didelon, P., Motte, F. et al. 2012, A\&A, 542, A81
\bibitem[Miville-Desch\^enes et al.(2010)]{Miville10} Miville-Desch\^enes, M.-A., Martin, P.G., Abergel, A. et al. 2010, A\&A, 518, L104
\bibitem[Miyama et al.(1997)]{Miyama97} Miyama, S.M., Narita, S., \& Hayashi, C.1987, Prog. Theor. Phys., 78, 1273
\bibitem[Molinari et al.(2010)]{Molinari10} Molinari, S., Swinyard, B., Bally, J. et al. 2010, {A\&A}, 518, L100
\bibitem[Motte et al.(1998)]{Man98} Motte, F., Andr\'e, P., \& Neri, R. 1998, A\&A, 336, 150
\bibitem[Motte et al.(2010)]{Motte10} Motte, F., Zavagno, A., Bontemps, S. et al. 2010, A\&A, 518, L77
\bibitem[Myers(1983)]{Myers83} Myers, P. C. 1983, ApJ, 270, 105
\bibitem[Myers(2009)]{Myers09} Myers, P.C.  2009, ApJ, 700, 1609
\bibitem[Nagasawa(1987)]{Nagasawa87} Nagasawa, M.  1987, Prog. Theor. Phys., 77, 635
\bibitem[Nakamura(1998)]{Nakamura98} {Nakamura}, F. 1998, ApJ, 507, L165
\bibitem[Onishi et al.(1998)]{onishi98} Onishi, T., Mizuno, A., Kawamura, A. et al. 
1998, ApJ, 502, 296
\bibitem[{{Ostriker}(1964)}]{Ostriker1964} {Ostriker}, J. 1964, ApJ, 140, 1056
\bibitem[Padoan \& Nordlund(2002)]{padoan02} Padoan, P. \& Nordlund, A. 2002, ApJ, 576, 870
\bibitem[Padoan et al.(2001)]{Pad01} Padoan, P., Juvela, M., Goodman, A.A., \& Nordlund, A. 2001, ApJ, 553, 227
\bibitem[Palmeirim et al.(2013)]{Palmeirim13} Palmeirim, P., Andr\'e, Ph., Kirk, J. et al. 2013, A\&A, 550, A38
\bibitem[Peretto et al.(2006)]{Peretto06} Peretto, N., Andr{\' e}, P., \& Belloche, A.\ 2006, A\&A, 445, 979 
\bibitem[Peretto et al.(2012)]{Peretto12} Peretto, N., Andr{\' e}, Ph., K\"onyves, V. et al.\ 2012, A\&A, 541, A63
\bibitem[Perrot \& Grenier(2003)]{Perrot03} Perrot, C.A., \& Grenier, I. A. 2003, A\&A, 404, 519
\bibitem[Pezzuto et al.(2012)]{Pezzuto12} Pezzuto, S., Elia, D., Schisano, E. et al.\ 2012, A\&A, 547, A54
\bibitem[Pilbratt et al.(2010)]{Pilbratt10} Pilbratt, G.L., Riedinger, J.R., Passvogel, T. et al. 2010, A\&A, 518, L1
\bibitem[Pon et al.(2011)]{Pon11} Pon, A., Johnstone, D., \& Heitsch, F. 2011, ApJ, 740, 88
\bibitem[Saigo \& Tomisaka(2011)]{Saigo11} Saigo, K. \& Tomisaka, K. 2011, ApJ, 728, 78
\bibitem[Schneider et al.(2013)]{Schneider13} {Schneider}, N., {Andr\'e}, Ph., {K\"onyves}, V. et al. 2013, ApJL, 766, L17
\bibitem[Schneider et al.(2010)]{Schneider10} {Schneider}, N., {Csengeri}, T., {Bontemps}, S. et al. 
2010, A\&A, 520, A49
\bibitem[Schneider \& Elmegreen(1979)]{Schneider79} Schneider, S. \& Elmegreen, B.G. 1979, ApJS, 41, 87
\bibitem[Shu(1977)]{Shu77} Shu, F. 1977, ApJ, 214, 488
\bibitem[Sousbie(2011)]{Sousbie11} Sousbie, T.,  2011, MNRAS,  414, 350
\bibitem[Stanke et al.(2006)]{Stanke06} Stanke, T., Smith, M. D., Gredel, R., \& Khanzadyan, T.\ 2006, A\&A, 447, 609
\bibitem[Starck et al.(2003)]{Starck03} Starck, J. L., Donoho, D. L., Cand\`es, E. J.\ 2003, A\&A, 398, 785
\bibitem[Toal\'a et al.(2012)]{Toala12} Toal\'a, J. A., V\'azquez-Semadeni, E., \& G\'omez, G. C.\ 2012, ApJ, 744, 190
\bibitem[Tohline(1982)]{Tohline82} Tohline, J.E. 1982, Fund. of Cos. Phys., 8, 1
\bibitem[Ward-Thompson et al.(2007)]{Ward07} Ward-Thompson, D., Andr\' e, P., Crutcher, R., Johnstone, D., Onishi, T., \& Wilson, 
C. 2007, Protostars and Planets V, Eds. B. Reipurth, D. Jewitt, K. Keil (Tucson: University of Arizona Press), p. 33
\bibitem[Ward-Thompson et al.(2010)]{Ward10} Ward-Thompson, D., Kirk, J.M., Andr\'e, P. et al. 2010, A\&A, 518, L92
\bibitem[Whitworth et al.(1996)]{Whitworth96} Whitworth, A.P., Bhattal, A.S., Francis, N., \& Watkins, S.J. 1996, MNRAS, 283, 1061 

\end{thebibliography}
\end{document}